%% file: main.tex
\documentclass[conference]{IEEEtran}
\pdfoutput=1

\usepackage{verbatim}
\usepackage{listings}
\usepackage{hyperref}
\usepackage[utf8]{inputenc}
\usepackage[T1]{fontenc}
\usepackage{cite}
\usepackage{mdframed}
\usepackage{enumitem}
\newmdtheoremenv{pitfall}{P}

\usepackage[dvipsnames]{xcolor}

\usepackage{lipsum}
\usepackage{cite}
\usepackage{tikz}
\usetikzlibrary{positioning}
\usetikzlibrary{fit}
\newcommand*\circled[1]{\tikz[baseline=(char.base)]{\node[shape=circle,draw, solid,inner sep=0.5pt] (char) {#1};}}
\usepackage{pgfplots}
\pgfplotsset{width=\columnwidth,compat=1.9}
\usepackage{filecontents} 
\usetikzlibrary{patterns,}

\newcounter{limitation}
\newenvironment{limitation}[1][]{\refstepcounter{limitation}\par\medskip\noindent
   \textbf{Limitation~\thelimitation---#1.} \rmfamily}{\medskip}

\usepackage{framed}
\newcounter{solution}
\newenvironment{solution}[1][]{\refstepcounter{solution}\par\medskip
   \begin{framed}
   \noindent\textbf{Takeaway \thesolution~--~#1.} \rmfamily
   } {
    \end{framed}
   }{\medskip}

\usepackage{marvosym}
\usepackage{amssymb}
\usepackage{pifont}
\usepackage{balance}
\usepackage{multirow}
\usepackage{array}
\newcolumntype{H}{>{\setbox0=\hbox\bgroup}c<{\egroup}@{ }}
\newcolumntype{P}[1]{>{\centering\arraybackslash}p{#1}}

\usepackage{mdwlist}
\usepackage{mathtools}
\usepackage{float}

\begin{document}

\title{The Forking Way: When TEEs Meet Consensus}

\author{\IEEEauthorblockN{Annika Wilde}
		\IEEEauthorblockA{Ruhr University Bochum\\
			annika.wilde@rub.de}
		\and
		\IEEEauthorblockN{Tim Niklas Gruel}
		\IEEEauthorblockA{Ruhr University Bochum\\
			tim.gruel@rub.de}
		\and
		\IEEEauthorblockN{Claudio Soriente}
		\IEEEauthorblockA{NEC Laboratories Europe\\
			claudio.soriente@neclab.eu}
		\and
		\IEEEauthorblockN{Ghassan Karame}
		\IEEEauthorblockA{Ruhr University Bochum\\
			ghassan@karame.org}}

\maketitle

\begin{abstract}
An increasing number of distributed platforms combine Trusted Execution Environments (TEEs) with blockchains. 
Indeed, many hail the combination of TEEs and blockchains a good ``marriage'': TEEs bring confidential computing to the blockchain while the consensus layer could help defend TEEs from forking attacks.

In this paper, we systemize how current blockchain solutions integrate TEEs and to what extent they are secure against forking attacks. To do so, we thoroughly analyze 29 proposals for TEE-based blockchains, ranging from academic proposals to production-ready platforms. We uncover a lack of consensus in the community on how to combine TEEs and blockchains. In particular, we identify four broad means to interconnect TEEs with consensus, analyze their limitations, and discuss possible remedies. Our analysis also reveals previously undocumented forking attacks on three production-ready TEE-based blockchains: Ten, Phala, and the Secret Network. We leverage our analysis to propose effective countermeasures against those vulnerabilities; we responsibly disclosed our findings to the developers of each affected platform.
\end{abstract}

\IEEEpeerreviewmaketitle

\section{Introduction}
\label{sec:intro}
\input{introduction}

\section{Background and Related Work}
\label{sec:preliminaries}
\input{background}

\input{blockchain_cloning}

\section{Case Studies}
\label{sec:case_studies}
In this section, we discuss in detail three new cloning attacks on three prominent production-ready TEE-based blockchains: Ten, Phala, and the Secret Network. These case studies serve to exemplify the pitfalls and mitigations discussed in Section~\ref{sec:bc_clon}. 

\begin{alignat*}{4}
&\textrm{Case Study 1}&&\xrightarrow[]{\text{uses}} &&\textrm{Serialization (\ref{sec:s_serial})}, \xrightarrow[]{\text{has}} &&\textrm{Limitations~\ref{lim:honesty},\ref{lim:forks},\ref{lim:random}} \\
&\textrm{Case Study 2}&&\xrightarrow[]{\text{uses}} &&\textrm{Stateless Enc. (\ref{sec:s_stateless})}, 
\xrightarrow[]{\text{has}} && \textrm{Limitations~\ref{lim:key_management},\ref{lim:throughput}-\ref{lim:forks}} \\
&\textrm{Case Study 3}&&\xrightarrow[]{\text{uses}} &&\textrm{Serialization (\ref{sec:s_serial})}, \xrightarrow[]{\text{has}} && \textrm{Limitation~\ref{lim:key_management}}
\end{alignat*}

We note that all of these blockchains share some design principles. Namely, they all rely on a global secret---shared among all enclaves---to derive cryptographic material,  such as encryption keys. For example, enclaves of the Secret Network share a so-called ``consensus seed'' used to derive, e.g., the public key used by a user to send encrypted requests to the enclave and the corresponding secret key used by enclaves to decrypt those requests. 
Another common design principle is the reliance on contract queries to save gas and improve latency. Phala and the Secret Network allow users to send contract queries to the enclaves via an HTTP endpoint. Such queries are encrypted using a public key derived from the network-wide secret. Our analysis considers rollback attacks on the sealed global secret out of scope. 

We responsibly disclosed our findings and recommendations to the Secret Network, Ten, and Phala, respectively, and shared our suggested countermeasures with their teams.

\subsection{Case Study 1: Phala}
\label{sec:cs_phala}
\input{phala}

\subsection{Case Study 2: The Secret Network}
\label{sec:cs_secret}
\input{secret}

\subsection{Case Study 3: Ten}
\label{sec:cs_ten}
\input{ten}

\section{Conclusion}
In this work, we provided a systemization of how current TEE-based blockchains resist forking attacks. To this end, we analyzed 29 TEE-based blockchains and showed an apparent lack of consensus in the community on how to leverage properties from distributed protocols to prevent forking attacks against TEE-based smart contracts.
More precisely, we showed that currently used mitigations for forking attacks introduce trade-offs in types of applications that can be deployed,  tolerance to peers joining/leaving the network, and overall complexity of the platform. 

Our study also revealed new forking vulnerabilities in three production-ready TEE-based blockchains: Phala, the Secret Network, and Ten.
We proposed effective countermeasures for each of those vulnerabilities, leveraging the results from our aforementioned analysis. We also responsibly disclosed our findings to the developers of each affected platform.

\section*{Acknowledgment}

This work is partly funded by the Deutsche Forschungsgemeinschaft (DFG, German Research Foundation) under Germany’s Excellence Strategy - EXC 2092 CASA - 390781972, and the European Union’s Horizon 2020 research and innovation program (REWIRE, Grant Agreement No. 101070627, and ACROSS, Grant Agreement No. 101097122). 
Views and opinions expressed are, however, those of the authors only and do not necessarily reflect those of the European Union. Neither the European Union nor the granting authority can be held responsible for them.

\bibliographystyle{IEEEtran}
\bibliography{references}

\end{document}

%% file: introduction.tex
Modern blockchains leverage smart contracts to run arbitrary business logic. 
Smart contracts instantiate state machine replication where all miners are expected to execute the contract code over common inputs and update their state. Here, the (binary) code of the smart contract and the transaction data must be available to all miners. In the case of applications handling sensitive data, this can hardly be tolerated.

Researchers and practitioners tried to address this gap by protecting the contract logic and the corresponding transaction data/state. Available solutions rely on trusted third parties to execute the contracts~\cite{bite,matetic2017rote}, zk-rollups~\cite{zkswap,zksync,aztec}, secure multi-party computation (MPC)~\cite{MPC_fabric,mpc_enigma}, or Trusted Execution Environments (TEEs)~\cite{fastkitten,whitepaper_phala,integritee,ekiden}. TEE-based solutions emerge as an attractive means to ensure the confidentiality of the contract and the associated data. Namely, they are (1) more efficient and more expressive compared to solutions based on MPC and (2) require drastically lower deployment costs compared to solutions that require trusted third parties. These factors led to considerable adoption of TEEs within existing decentralized platforms, such as Ten~\cite{ten_website}, Phala~\cite{whitepaper_phala}, and the Secret Network~\cite{graypaper_secret_network}. Given the pervasiveness of TEEs nowadays, the number of decentralized platforms supporting TEEs is only expected to grow.

While TEEs bring several benefits to blockchains (e.g., confidential computing for smart contracts), they can also leverage the consistency layer of the underlying blockchain to mitigate one of their fundamental limitations: the lack of proper countermeasures against so-called \emph{forking attacks}~\cite{cachin_collective_mem}. 
Such attacks can be mitigated if the TEE processes requests that are properly serialized. Consensus protocols, in general, and blockchains, in particular, are notorious for ensuring the total ordering of events. Hence, TEE-based applications can naturally rely on blockchains to counter forking attacks~\cite{sgx_oneratete_paper, li2022sokteeassistedconfidentialsmart}. On an abstract level, such a ``marriage'' between blockchains and TEEs supports the confidential execution of contract logic while mitigating forking attacks on TEEs. 

In this paper, we provide a systematization of existing solutions used by various TEE-based blockchains to counter forking attacks against the enclave.  
We point out that a forking attack against TEEs can be carried out either by rolling back its state or by cloning the TEE instance~\cite{cachin_collective_mem}. 
Unfortunately, while previous work~\cite{sgx_oneratete_paper, li2022sokteeassistedconfidentialsmart} investigated forking attacks based on rollbacks, it did not consider practical forking threats that can arise from cloning. In other words, a TEE-based blockchain that includes anti-rollback mechanisms may still be susceptible to forking attacks based on cloning.
Towards this end, we analyze 29 proposals for TEE-based blockchains, ranging from academic proposals to production-ready platforms.
We identify four broad categories of anti-forking techniques used in TEE-based blockchains.
Further, we analyze the trade-offs of each technique---ranging from the expressiveness of the smart contract that can be deployed in the TEE to the restriction on the L1 layer that can be used. We then highlight pitfalls in how these techniques are currently instantiated and discuss workable mechanisms to practically address those pitfalls. 
Our analysis shows that combining TEEs and blockchains to provide forking-resistant confidential smart contracts presents a number of practical challenges that are often overlooked by researchers and practitioners. 
In particular, we show that (1) stateless enclaves can be protected against forking attacks in existing protocols by leveraging ephemeral enclave identities, but (2) devising comprehensive solutions to protecting stateful enclaves in existing TEE-based blockchains depends on several factors, such as the type of consensus (final or eventual) or the throughput of the consensus layer, among others.

Throughout our systematic analysis, we identify several vulnerabilities that lead to forks in the TEE state.
Among these vulnerabilities, we describe previously undocumented forking attacks on three production-ready TEE-based blockchains, namely: Ten~\cite{ten_website}, Phala~\cite{whitepaper_phala}, and the Secret Network~\cite{graypaper_secret_network}. 

In summary, we make the following contributions:
\begin{basedescript}{\desclabelstyle{\pushlabel}}
\item[Systemization of knowledge: ] We provide a systemization of existing solutions used to counter forking attacks on TEE-based blockchains. We analyze, by means of a thorough empirical analysis,  29 proposals for TEE-based blockchains. We categorize existing anti-forking mechanisms used by such platforms into four broad categories and identify their gaps and limitations (cf. Section~\ref{sec:bc_clon}).
\item[Pitfalls in TEE-based blockchains: ] We explore the solution space to secure TEE-based blockchains against possible forking attacks. Namely, we discuss the various trade-offs existing solutions exhibit and show significant pitfalls
in how these techniques are instantiated. For example, Li \emph{et al.}~\cite{li2022sokteeassistedconfidentialsmart} note that FastKitten~\cite{fastkitten} cannot be forked by rollback. However, we show a new cloning-based forking attack against FastKitten (cf. Section~\ref{sec:s_permissioned}). 
We discuss workable mechanisms to address those pitfalls practically and show that the underlying choice of the mitigation technique depends mostly on the provisions of Layer 1. 
\item[Cloning vulnerabilities in production-ready networks:] We present and evaluate previously undocumented cloning vulnerabilities in Ten, Phala, and the Secret Network, three production-ready TEE-based blockchains (cf. Section~\ref{sec:case_studies}). Our first attack against Phala enables an adversary to operate two
instances of the enclave contract (with the same code and address) and freely choose the instance to answer a request; given that the two instances share the code and the address, a client cannot distinguish which one answered its request. This allows the adversary to reply with a stale state and works despite defenses like transaction ordering and other well-known measures used to prevent rogue contract injections~\cite{rogue_contract_blog}. 
Similarly, in our second attack against the Secret Network, the adversary uses another instance of the same contract code to answer the query incorrectly based on a different state. 
This new attack works despite the anti-rollback measures proposed by Jean-Louis \emph{et al.}~\cite{sgx_oneratete_paper}.  Our final novel attack against Ten allows an adversary to spawn as many enclaves to 
increase the chances that it is elected as the next rollup proposer despite the anti-rollback solution used in Ten.

\item[Practical countermeasures:] By leveraging our systematic analysis, we discuss and analyze practical and workable solutions, based on our findings, to address the vulnerabilities identified in Ten, Phala, and the Secret Network.
\end{basedescript}

\noindent \textbf{Responsible disclosure: } We responsibly disclosed our findings and suggested countermeasures to the 
developers of these production-ready TEE-based blockchains, respectively (see \url{https://cloning-tee-blockchains.github.io/}).

%% file: background.tex
\subsection{Hardware-based TEEs}

Trusted Execution Environments leverage the hardware to control access to runtime memory by software, thereby providing an isolated sandbox---known as ``enclave''---to execute user code. As such, the threat model for TEEs includes malicious user (peer) processes and a malicious OS while the underlying hardware is trusted. 
Commercial TEEs include Intel SGX~\cite{sgx_intro}, AMD SEV~\cite{sev_snp}, or ARM TrustZone~\cite{trustzone}. While each of those commercial TEEs has its own unique features, they all share a common blueprint. In the following, we will only discuss the TEE features that are relevant for this work and refer the reader to~\cite{sgx_explained} for a complete treatment of TEEs. 

\vspace{0.5em}\noindent \textbf{Attestation and Enclave Identity.} Attestation allows (remote) verifiers to check the code that is running within an enclave and the configuration of the underlying platform. In a nutshell, a trusted system component uses a private key to sign a hash computed over the code deployed in the enclave and various attributes of the machine (e.g., TEE version, security patches); the verifier uses a public key, usually distributed by the TEE manufacturer, to verify the signature. 

The hash computed over the code and the machine attributes is often referred to as the ``identity'' of the enclave and allows to distinguish two enclaves running two different binaries on the same machine or two enclaves running the same binary but on two different machines. Note that TEEs provide no support to distinguish enclaves with the same binary on the same platform.  

\vspace{0.5em}\noindent \textbf{Sealing.} Apart from secure runtime memory, TEEs also provide secure (disk) storage. This is achieved by means of authenticated encryption and by leveraging so-called \emph{sealing} keys. A sealing key for a given enclave is derived from a platform-specific master key and the identity of the enclave. Hence, two enclaves running different binaries (or on different platforms) cannot access the same sealing key; as a result, data sealed by one enclave cannot be unsealed by the other. Nevertheless, two enclaves on the same platform running the same binary have access to the same sealing key. 
We point out that access to the disk is mediated by a possibly malicious OS; hence, sealing provides no freshness guarantees. That is, when the enclave fetches a sealed state from the disk, it has no means to distinguish whether the ciphertext provided by the OS corresponds to (1) the latest sealed state or (2) to an older ciphertext containing a stale state.

\subsection{Forking Attacks on TEEs}
\label{sec:bg_forking}
Forking attacks are well-known threats to the consistency of distributed applications~\cite{long-range_attacks}. In the context of TEEs, a forking attack leverages 
the lack of freshness of the sealing functionality or the lack of mechanisms to distinguish two instances of the same enclave application. In other words, a forking attack on a TEE can be mounted either by rolling back the enclave state or cloning the enclave application. 

In the following, we describe both strategies with a sample TEE application denoted as $E$. Consider the case where $E$ updates its state $s_j$ based on a function $f$, the previous state $s_{j-1}$, and an input $i$. Let $s_0$ be the enclave starting state, and assume that $E$ seals its state to disk after every update to recover it in case of crashes. Furthermore, let $E$ receive two inputs, $i_1$ and $i_2$, one after the other. In a benign setting, the enclave will move through states $s_1 = f(s_0, i_1)$ and $s_2 = f(s_1, i_2)$, each sealed to disk. 

\vspace{0.5em}\noindent \textbf{Rollback-based forking.}
Forking can be achieved by terminating the enclave after it has sealed $s_1$. Next, $E$ is restarted; when the enclave fetches state from disk, the adversary provides $s_0$ instead of $s_1$, thereby rolling back the state of the enclave. If the enclave then processes $i_2$, it moves to state $s_2' = f(s_0, i_2)$.

\vspace{0.5em}\noindent \textbf{Cloning-based forking.} Here, the adversary launches another enclave instance running the same code---denoted as $E'$---on the same machine. Note that, as discussed above, it is not possible to distinguish the two instances apart. Hence, the adversary feeds $i_1$ to $E$ that advances to state $s_1=f(s_0,i_1)$ and feeds $i_2$ to $E'$ that advances to state $s'_1=f(s_0,i_2)$.

\subsection{Layer One, Layer Two, and Blockchain Applications}

A blockchain is a decentralized system that ensures state consistency among the system's nodes. Consistency is maintained through a ``Layer One'' (L1) consensus protocol, which ensures that the majority of (honest) nodes agree on a common state. Transactions are used to update the state, and nodes typically use the consensus protocol to agree on the order of transactions, organized in batches called blocks. For instance, Nakamoto-style blockchains (such as Bitcoin and Ethereum Classic) use a randomized leader election protocol (leveraging Proof of Work or Proof of Stake) to elect the next block proposer but rely on the longest chain rule to reach consensus on transactions and blocks.  
Such protocols scale well to a high number of nodes but only achieve modest throughput. 

Layer Two (L2) solutions overcome the performance and functionality limitations of L1. 
For example, payment channels~\cite{DBLP:conf/fc/MohantyT23} and roll-ups~\cite{DBLP:conf/aft/ShengRBTV24} enable the bulk processing of transactions, thereby decreasing the transaction processing load on L1. Similarly, cross-chain bridges~\cite{DBLP:conf/ccs/XieZCZ0JBS22} and atomic swap protocols~\cite{DBLP:conf/sp/ThyagarajanMM22} enable interaction between independent blockchains. These solutions, however, still rely on the underlying L1 blockchain for final confirmation and validation. For example, L2 solutions that process batches of transactions off-chain must periodically commit state updates to the L1 blockchain, which provides final confirmation for those off-chain transactions.

Some blockchain applications interface L1 blockchains with L2 solutions. For example, lightweight clients~\cite{gervais2014bloomFilters} monitor the blockchain by storing only block headers, querying transactions from full nodes, and verifying the responses using these headers. Other applications like wallets~\cite{gutoski2015bitcoinWallets} securely manage user accounts and the keys necessary to authorize transactions. 

\subsection{Related Work}

To the best of our knowledge, no previous work has focused on forking attacks against TEE-based blockchains. We now discuss the most relevant related papers. 

Li \emph{et al.}~\cite{li2022sokteeassistedconfidentialsmart} systematize TEE-assisted confidential smart contracts with respect to ``Privacy-Preserving Properties'' (e.g., I/O privacy) and  ``Blockchain Intrinsic Benefits'' (e.g., high availability). The authors of~\cite{li2022sokteeassistedconfidentialsmart} briefly mention ``state consistency'' and the problem of state freshness for TEEs. Nevertheless, Li \emph{et al.}~\cite{li2022sokteeassistedconfidentialsmart} do not provide a systematic study on forking vulnerabilities of TEE-based blockchains. 
Jean-Louis \emph{et al.}~\cite{sgx_oneratete_paper} investigate privacy flaws in four production-ready blockchains, namely Phala, Ten, Oasis, and the Secret Network. The authors discuss the privacy leaks due to a potentially malicious OS that sees enclave accesses to an external (encrypted) database or the enclave page faults. The authors of~\cite{sgx_oneratete_paper} also show a rollback attack on the Secret Network that allows the adversary to learn private information about transactions (e.g.,  transaction amounts, balances, etc.). However, Jean-Louis \emph{et al.}~\cite{sgx_oneratete_paper} provide no attacks on the other platforms they consider, nor do they take into account attacks based on cloning the enclave.

%% file: blockchain_cloning.tex
\section{Systemization Methodology}

\begin{figure*}
    \centering
	\includegraphics[width=1.0\textwidth]{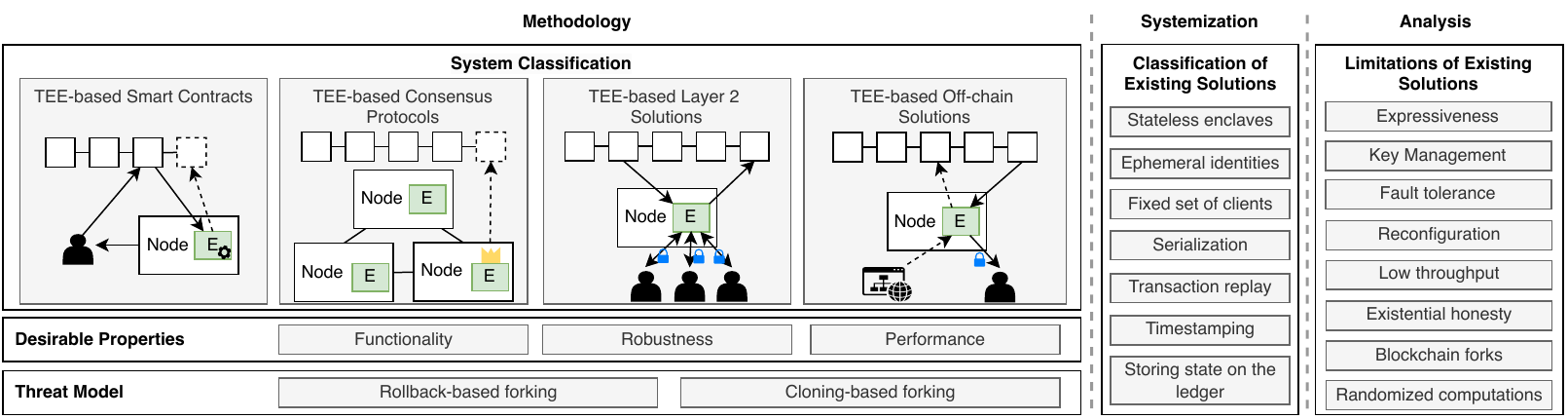}
    \caption{Overview of our systemization methodology: We classify SGX-based blockchains into four distinct categories and analyze their tradeoffs with respect to robustness, functionality, and performance.}
    \label{fig:systemization}
\end{figure*}

In what follows, we explain our classification criteria and outline our systemization methodology. Figure~\ref{fig:systemization} provides an overview of our approach.

\subsection{Selection Criteria }
To the best of our knowledge, the majority of TEE-based blockchains rely on Intel SGX, with two notable exceptions: CCF~\cite{howard2023ccf}, which supports both Intel SGX and AMD SEV, and TZ4Fabric~\cite{muller2020tz4fabric}, which is based on ARM TrustZone. Consequently, our systematic evaluation focused solely on SGX-based blockchains. However, we believe our findings are applicable to other TEEs, as they share similar limitations.

The selected platforms were taken from a curated list of SGX-based blockchains~\cite[Sec. Blockchains/Session 4]{awesome_sgx} and an SoK on TEE-assisted confidential smart contracts~\cite{li2022sokteeassistedconfidentialsmart}, totaling 41 platforms.
We discarded platforms without an enclave instantiation, platforms that were archived, or that had no English documentation, ending up with 28 platforms. We considered an additional production-ready platform,  Ten~\cite{ten_website}, that was studied by~\cite{sgx_oneratete_paper}---one of the closest related works.

\subsection{System Classification}
\label{sec:sys_classification}

We systemize the platforms in our analysis based on how they leverage functionality from TEEs. We  identify four main categories:

\begin{description}[leftmargin =*, itemsep=0.5 em]
    \item[Category 1---TEE-based Smart Contracts: ] 
    Blockchains in this category use the TEE to achieve confidentiality in the execution of the smart contract. That is, the TEE fetches encrypted inputs from L1, processes the transactions, and pushes encrypted outputs to L1. 
    Some blockchains~\cite{howard2023ccf,yan2020confide} employ state-machine replication, where the smart contract state is replicated across multiple enclaves, whereas others~\cite{whitepaper_phala} assign a single enclave per contract and verify its integrity via attestation.  The contract's state may be stored locally within the enclave~\cite{yan2020confide,graypaper_secret_network,howard2023ccf}, or on the blockchain itself~\cite{ekiden}. To prevent a malicious smart contract from accessing the state of another (a scenario known as rogue smart contract code injection~\cite{rogue_contract_blog}), the enclave typically binds the contract address (derived from the code hash) with the state. 
    Note that the blockchain may allow clients to issue \emph{contract queries}, i.e., direct read requests to the smart contract that bypass the ledger and allow the client to obtain information on the contract state~\cite{graypaper_secret_network,whitepaper_phala}. 
    \item[Category 2---TEE-based Consensus Protocols:] 
    Blockchains in this category use the TEE to speed up or scale consensus. 
    Some blockchains~\cite{rem_pouw,proof_of_luck,engraft} leverage the TEE for secure leader election. For example, the ``Proof of Luck''~\cite{proof_of_luck} consensus protocol uses the TEE as a source of unbiased randomness to select the next block proposer. Other systems~\cite{howard2023ccf,veronese2013minBFTminZyzzyva} execute the consensus mechanism directly within the TEE to improve scalability. Since the code inside the TEE is assumed to be trusted after attestation, this approach dramatically reduces communication complexity in the presence of malicious nodes.

    \item[Category 3---TEE-based L2 Solutions:] 
    Blockchains in this category leverage the TEE to implement confidential L2 solutions. 
    Some blockchains~\cite{fastkitten,private_chaincode,pdo,integritee} use the TEE to instantiate confidential smart contracts for L1 blockchains that lack native support for such features. 
    Other L2 solutions focus on supporting confidential operations over transactions, including mixers~\cite{obscuro_mixer}, payment channels~\cite{lind2019teechain,twilight,erwig2023commitee}, and cross-chain bridges~\cite{bentov2019tesseract}.
    
    \item[Category 4---TEE-based  Blockchain Applications: ] Last but not least, some applications leverage TEEs to enable secure access to the blockchain. This includes the secure storage of key material required for blockchain interaction~\cite{sgxwallet,doc_ternoa}, secure fetching and validation of blocks~\cite{bite}, and secure data retrieval for blockchain-based applications~\cite{town_crier}.    
\end{description}

\subsection{Methodology}

We analyze the selected platforms to identify the various techniques used to prevent forking attacks. 
We detect four broad techniques. Some platforms use \emph{stateless enclaves} that cannot be rolled back ``by design'' but may be vulnerable to cloning. Other platforms rely on \emph{ephemeral enclave IDs} to distinguish clones. Another technique is to rely on a \emph{fixed set of clients} that monitors the enclave to detect forks. Finally, a common alternative anti-forking technique is to \emph{serialize the enclave state} by using the ledger. 

These techniques, however, incur several trade-offs in terms of \emph{functionality}, \emph{robustness}, and \emph{performance}. For example, stateless enclaves cannot run stateful applications, thereby restricting the functionality offered by the smart contract. Similarly, using ephemeral IDs or a fixed set of clients hinders identity management and complicates the addition/removal of nodes. Finally, using the ledger to serialize the enclave state bounds the enclave throughput to the throughput of the ledger, thereby reducing performance. 

In the following sections, we systematize the selected 29 platforms based on the anti-forking mechanisms they use, assess their resistance to forking attacks, and analyze their robustness, performance, and functionality trade-offs.

\section{Pitfalls in TEE-based Blockchains}
\label{sec:bc_clon}

We assume the typical threat model for SGX applications, where the hardware is part of the trusted computing base (TCB), but any privileged software, such as the operating system (OS), is considered potentially malicious. In this model, the adversary fully controls system resources, including memory, storage, and network communication, but cannot compromise the hardware.
Our analysis focuses explicitly on forking attacks within the context of TEE-based blockchains. As such, we assume that the enclave remains uncompromised and that side-channel and denial-of-service (DoS) attacks are out of scope. 
Whenever applicable, we assume that the adversary controls a node running the enclave. Since the adversary controls the communication between the enclave and the rest of the system, they can drop or modify all inputs and outputs of the enclave. For instance, the adversary can provide the enclave with a stale state whenever the enclave fetches the state from the disk or the blockchain. Further, the adversary can clone the enclave by launching an arbitrary number of instances of the same TEE binary. As mentioned in Section~\ref{sec:preliminaries}, SGX provides no means to distinguish two enclaves running the same binary on the same platform, and the two binaries can access the same sealed state. 

\subsection{Stateless Enclaves}
\label{sec:s_stateless}

\noindent\textbf{Overview.} 
A stateless enclave produces output depending only on the current input and does not need to maintain the state of previous computations. Hence, if restarted, the enclave fetches no state from persistent storage~\cite{russinovich2019ccf,obscuro_mixer,engraft,lskv_talk,mobilecoin,whitepaper_phala,pdo,proof_of_luck,rem_pouw,graypaper_secret_network,sgxwallet,bentov2019tesseract,town_crier,twilight,erwig2023commitee,xiao2020privacyGuard}. 
At times, the enclave may use an immutable state, such as a signing key, fetched from persistent storage upon every restart. However, the state (i.e., the signing key) never changes.
Prominent examples of stateless enclaves are transaction mixers~\cite{obscuro_mixer} that output a permutation of the set of transactions received as input.   Differently, an enclave that implements a standard database is typically not stateless---since previous queries may determine the result of the current query---and the database is periodically sealed to disk to be fetched upon restarts.

\vspace{0.5em} \noindent \textbf{Vulnerability to Forking Attacks: }
Stateless enclaves are resistant to rollback attacks \emph{by design}. If the enclave fetches no state from persistent storage or fetches an immutable state, then the adversary has no means to roll back the enclave. Nevertheless, cloning attacks remain viable in this setting. For example, if the computation is randomized, the adversary can launch multiple clones and select the more favorable output.
    
\begin{limitation}[Expressiveness] 
\label{lim:stateless_1}
    A stateless enclave clearly limits the type of applications that can be deployed in the smart contract. For example, a database application requires enclaves to keep a persistent state and cannot be deployed in a smart contract running within a stateless enclave. 
\end{limitation}

\noindent\textbf{Example: PoUW.} Zhang \emph{et al.}~\cite{rem_pouw} propose a leader election protocol based on Proof-of-Useful-Work (PoUW). PoUW is an alternative to well-known Proof-of-Work protocols where miners ``waste'' resources to solve a cryptographic puzzle. In PoUW, mining resources are used to carry out useful tasks submitted by clients. A miner can propose a new block only if it produces ``proof'' that it has completed a specific task.  
In particular, after completion of a task, the miner draws a random number $r \in [0,1]$ and proposes a new block only if $r$ is greater than a threshold $t$ set to $1 - (1 - \mathrm{diff})^n$, where $\mathrm{diff}$ is a tuneable difficulty parameter, and $n$ is the number of instructions required to complete the task---so that tasks with more instructions increase the chances of the miner to propose a block. The proof is, essentially, the task output $out$ and the random number $r$.
PoUW leverages TEEs to ensure that the miner faithfully completes the task and draws an unbiased random $r$. In particular, the task is executed within an enclave; upon completing the task, the enclave draws $r$ and signs the proof with its signing key. Other nodes can verify the proof using the corresponding public key. Remote attestation allows nodes to check that the miner indeed runs a legitimate PoUW enclave. Note that enclave signatures also include the current block's hash so that a PoUW has limited validity. This measure prevents rollback attacks: a malicious host may feed (the hash of) a stale block to its PoUW enclave, but the proof that the enclave outputs will not be accepted by other nodes---since it is not tied to the current block. 

\begin{figure}[]
    \centering
    \scalebox{0.75}{
        \includegraphics[]{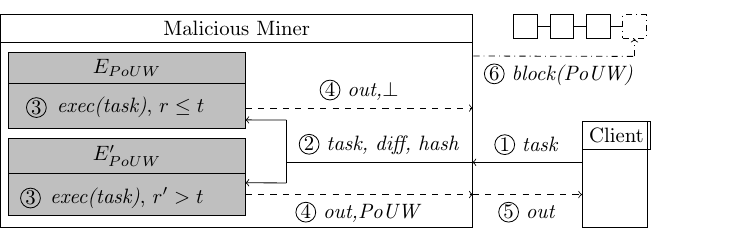}
    }
    \caption{Example of a cloning attack on PoUW~\protect\cite{rem_pouw}. A malicious miner starts two PoUW enclaves to increase its chances that one of its enclaves produces proof of useful work.}
    \label{fig:cloning_rem}
\end{figure}

Despite being secure against rollback attacks, a cloning attack on PoUW is still possible, as shown in Figure \ref{fig:cloning_rem}. Assume a malicious miner that receives a \emph{task} from a client (step \circled{1}). The miner starts two PoUW enclaves $E_{PoUW}$ and $E'_{PoUW}$ and provides them with the received task, the current block, and difficulty (step \circled{2}). The enclaves execute the task, yielding the same $out$ and $n$, so they compute the same threshold $t$.  $E_{PoUW}$ and $E'_{PoUW}$ now draw random numbers $r$ and $r'$, respectively. Assume that $r'>t$ while $r\leq t$ (step \circled{3}). Consequently, $E'_{PoUW}$ returns a PoUW, while $E_{PoUW}$ does not (step \circled{4}). The adversary returns the output to the client (step \circled{5}) and proposes a new block with the PoUW generated by $E'_{PoUW}$ (step \circled{6}). The adversary effectively increased its chances of proposing the next block by running two
clones of the PoUW enclave. The adversary may run $>2$ PoUW enclaves to increase its chances of proposing the next block further.

\begin{solution}[Stateless Enclaves]
\label{sol:stateless}
    Using enclaves that do not keep a persistent state protects against rollback attacks by design. However, stateless enclaves limit the expressiveness of the TEE application and do not deter cloning attacks when the TEE application is non-deterministic.
\end{solution}

\subsection{Ephemeral Identities}
\label{sec:s_ephemeral}

\noindent\textbf{Overview.} 
Most TEE-based blockchains use enclaves with long-lasting identities~\cite{ten_website,mobilecoin,engraft,town_crier,lind2019teechain,private_chaincode}. For example, one can identify an enclave with its public key, and the key pair is sealed to disk to recover from crashes. However, some blockchains assign ephemeral IDs to enclaves~\cite{obscuro_mixer,engraft,doc_ternoa,bentov2019tesseract,twilight,erwig2023commitee,wang2020hybridchain}. That is, the enclave generates a fresh key pair at runtime, and the public key is used as an identifier. The key pair is not sealed to the disk. Hence, if the enclave crashes and is restarted, it obtains a new identity.

\vspace{0.5em} \noindent \textbf{Vulnerability to Forking Attacks: } This simple technique prevents cloning attacks since ephemeral public keys allow external parties to distinguish two enclave instances, even if they run the same binary on the same platform---since they will likely generate two different key pairs. Hence, messages encrypted under the ephemeral public key of one enclave instance cannot be decrypted by any other instance. Note that if the enclave seals the state to disk, this solution does not protect against rollback attacks. We note that all the applications we analyzed and used ephemeral keys have stateless enclaves.

\begin{limitation}[Key management]
\label{lim:key_management}
    Ephemeral keys as an anti-cloning mechanism may require proper key management. That is, the consensus protocol must keep track of participating enclaves and their (ephemeral) public keys. For example, in the case of PoUW~\cite{rem_pouw}, an ephemeral key per enclave would prevent the adversary from using several instances of the PoUW enclave as long as there is a consistent layer that keeps track of all registered ephemeral public keys. This requires a dedicated registration process. Further, when an enclave crashes, a mechanism is needed to remove the old ephemeral key from the list of participating enclaves and add the new one (created by the enclave upon restart). 
    
    In case the application logic can tolerate multiple clones of an enclave without causing harm, there is no need to rely on a dedicated key management mechanism.
    For instance, Tesseract~\cite{bentov2019tesseract} allows clients to send time-locked deposits by encrypting coins under the ephemeral public key of the Tesseract enclave. The enclave does not seal state, and, in case of a crash, coins are automatically reverted to the original client account after the time-lock expires. Thus, the Tesseract enclave logic is inherently robust to cloning attacks; that is, an adversary gains no advantage by cloning the Tesseract enclave.
\end{limitation}

\noindent\textbf{Example: Twilight.} Dotan \emph{et al.}~\cite{twilight} propose Twilight, a differentially private payment channel network. Payment channels are established between two relays, and assets can be transferred between them, bypassing the blockchain. Two parties that are not connected via a payment channel can transfer coins using a sequence of hops between relays.
At startup, the Twilight enclave generates an ephemeral key pair. Other relays use the public key of a relay to send encrypted payment information. Note that Twilight enclaves do not store state on disk. Thus, even if they cannot tolerate crashes, they resist rollback attacks by design (cf. Section~\ref{sec:s_stateless}).

\begin{figure}[]
    \centering
    \scalebox{0.75}{
        \includegraphics[]{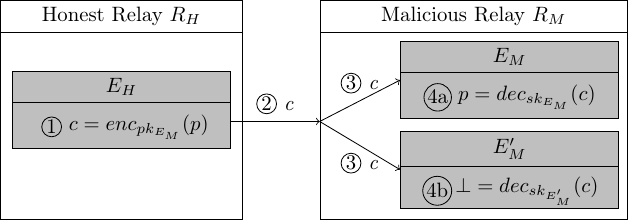}
    }
    \caption{Example of a (failed) cloning attack on Twilight~\protect\cite{twilight}. A malicious relay forwards an encrypted payment to two instances of the Twilight enclave. Ephemeral keys prevent enclave $E'_M$ from decrypting the ciphertext.
    }
    \label{fig:cloning_twilight}
\end{figure}

Figure~\ref{fig:cloning_twilight} shows how Twilight leverages ephemeral keys to prevent cloning attacks. Assume an honest relay $R_H$ and a malicious relay $R_M$. $R_H$ operates Twilight enclave $E_H$, whereas $R_M$ operates two enclaves $E_M$ and $E'_M$. Each enclave generates a key pair. $E_H$ and $E_M$ exchanged public keys to establish a payment channel. A dedicated smart contract binds the payment channel to the ephemeral keys of $E_H$ and $E_M$ (addressing Limitation~\ref{lim:key_management}). Assume that $E_H$ wants to send payment $p$ over this channel and encrypts it with $pk_{E_M}$, resulting in ciphertext $c$ (step \circled{1}). $E_H$ outputs $c$ to $R_H$ which sends it to $R_M$ (step \circled{2}). $R_M$ tries to claim the payment twice, sending it to both enclave instances (step \circled{3}). $E_M$  decrypts $c$ and retrieves $p$ because it has the secret key corresponding  to $pk_{E_M}$ (step \circled{4a}). However, $E'_M$ has a different secret key; hence decryption of $c$ fails (step \circled{4b}).

\begin{solution}[Ephemeral Identities]
\label{sol:eid}
    Identifying each enclave by means of an ephemeral ID (i.e., renewed at restart) can prevent cloning attacks. In settings where the state needs to persist, one should additionally rely on anti-rollback mechanisms.
\end{solution}

\subsection{Fixed Set of Clients}
\label{sec:s_permissioned}

\noindent\textbf{Overview: } Some TEE-based blockchains only allow a fixed set of clients to interact with the smart contract~\cite{fastkitten,lind2019teechain}. For example, FastKitten~\cite{fastkitten} implements a smart-contract solution on top of Bitcoin. Time is divided into rounds; at each round, all clients send their inputs and views of the (latest) contract state to the contract. The contract processes the inputs and moves to the next round only if its local state matches the views received by all clients. A similar approach is used in Lightweight Collective Memory (LCM)~\cite{cachin_collective_mem} where a set of mutually trusted clients interacts with a TEE application and exchange 
their view of the system to detect inconsistencies.

\vspace{0.5em} \noindent \textbf{Vulnerability to Forking Attacks: }
A fixed set of clients may help prevent forking attacks based on rollbacks. For example, a FastKitten smart contract can detect a rollback if the clients' state information does not match the local state. Similarly, in LCM, clients can detect if any response from the smart contract does not match the client's global view.
We note, however, that a solution with a fixed set of clients does not prevent cloning attacks: the adversary may still have an advantage in running multiple clones of the enclave. For example, when the computation is randomized, the
adversary can launch multiple clones and forward the more favorable output to the clients.

\begin{limitation}[Fault tolerance]
\label{lim:fault}
    This approach requires clients to be online and to trust each other. Hence, the application does not tolerate client crashes or byzantine faults.
\end{limitation}

\begin{limitation}[Reconfiguration]
\label{lim:reconfig}
    Reconfiguration operations, i.e., allowing clients to join or leave the set of participating clients, are costly operations that require re-negotiations of all established cryptographic keys in the system.
\end{limitation}

\vspace{0.5em} \noindent\textbf{Example: FastKitten.}  Das \emph{et al.}~\cite{fastkitten} propose a TEE-based solution to deploy smart contracts on Bitcoin. FastKitten consists of an enclave $E_Q$ that executes a smart contract with a fixed set of clients. During setup, all clients deposit coins to $E_Q$'s address, which are then redistributed during the contract execution. The execution of the smart contract is split into computation rounds. The enclave redistributes the coins in each round $j$, advancing from state $s_{j-1}$ to $s_{j}$. The enclave regularly seals the state to provide fault tolerance. At the beginning of a round, each client $i$ signs the previous state $s_{j-1}$ and its input $I_{i,j}$ for the current round.
If the enclave state is rolled back, the local state will not match the state sent by clients. Note, however, that \cite{fastkitten} does not address Limitation~\ref{lim:reconfig}. 

\begin{figure}[]
    \centering
    \scalebox{0.75}{
        \includegraphics[]{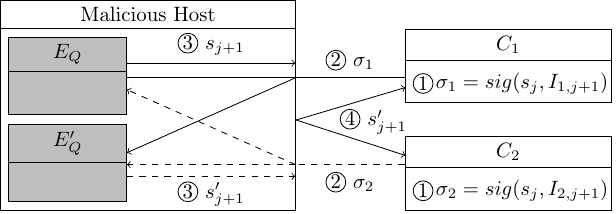}
    }
    \caption{Example of a cloning attack against a non-deterministic smart contract in FastKitten~\cite{fastkitten}. A malicious host starts two enclaves and selects the preferred output.}
    \label{fig:cloning_fastkitten}
\end{figure}

Despite being secure against rollback attacks, a cloning attack here is still possible if the enclave executes a probabilistic smart contract (e.g., a lottery contract as suggested in~\cite{fastkitten}). Assume a malicious host that runs an enclave $E_Q$ that has already completed the setup phase with two clients, as shown in Figure~\ref{fig:cloning_fastkitten}. Let the enclave state be $s_j$ and assume it is sealed to disk. The malicious host now starts a clone $E'_Q$ and provides it with $s_j$. In the next round, each client $i$ binds $s_j$ to its input $I_{i,j+1}$ by means of a signature (step~\circled{1}). The clients submit their signatures to the host that feeds it to both $E_Q$ and $E'_Q$ (step~\circled{2}). Both enclaves successfully verify the signatures and treat $s_j$ as the current valid state. At this stage, each enclave computes a different output since the lottery contract is randomized (step~\circled{3}). The adversary selects the most favorable output (e.g., the output that favors a specific client)  and forwards it to the clients (step~\circled{4}).

\begin{solution}[Fixed Set of Clients]
\label{sol:permissioned}{
    Relying on a fixed and mutually trusted set of smart contract clients can prevent rollback attacks; however, it cannot prevent cloning attacks if the enclave is non-deterministic.
    }
\end{solution}

\subsection{Serializing State}
\label{sec:s_serial}

\noindent\textbf{Overview.}
Most platforms use the blockchain's ordering layer to persist and serialize the enclave state. Across the platforms we analyzed, we witnessed three variants of this technique.

\vspace{0.5em} \noindent\textbf{Option 1---Transaction replay from the ledger.}
Some platforms do not use the local sealing functionality for the enclave to recover the state after a crash. Instead, the enclave recovers state by fetching all blocks from the blockchain and processing one block after another~\cite{russinovich2019ccf,lskv_talk,whitepaper_phala,bentov2019tesseract,graypaper_secret_network}. 
This option can prevent rollback attacks if the enclave obtains the complete set of blocks up to the current one. It can also protect against cloning if the blockchain cannot be forked. 

\vspace{0.5em} \noindent\textbf{Option 2---Timestamping.} 
Other platforms use sealing to persist state locally (e.g., on disk) but include block metadata---such as its height and hash---in their state as an anchor to ensure state freshness~\cite{russinovich2019ccf,bite,credb,crust_sworker,lskv_talk,whitepaper_phala,bentov2019tesseract,doc_ternoa,town_crier,yan2020confide,yuan2018shadowEth}. For instance, the block height---i.e., the number of blocks from the genesis block until the current block---can be used as a logical clock to track which transactions were committed to the ledger and which were processed by the enclave.  
Here, it is paramount that an enclave includes its current timestamp (e.g., the height of the last block it has processed) in its responses to contract queries. This burdens the requesting client to compare the timestamp in the ledger (i.e., the current block height) with the one included in the response from the enclave to detect forking attacks. 
Some platforms, such as TERNOA~\cite{doc_ternoa}, allow clients to specify a range of block heights for their contract queries (to cater to partial-synchronous deployments). Here, a query includes a minimum block height $m$ and a maximum block height $M$. The enclave serves the query only if the height of the latest processed block falls within $[m, M]$. This design choice requires clients to keep track of the current block height and only ensures loose synchronization between the ledger and the TEE. In particular, an adversary could roll back the enclave to a state where the latest block has not been processed yet to ensure that the TEE's answer does not take a recent transaction into account.
In other platforms, such as IntegriTEE~\cite{integritee}, the enclave regularly sends a heartbeat transaction to the blockchain, including the current block height. The enclave only answers contract queries if it receives the corresponding acknowledgment. 

\vspace{0.5em} \noindent\textbf{Option 3---Storing state in the blockchain.} 
Another variant that we witnessed involves enclaves that seal their state (representation) on the ledger~\cite{ekiden,engraft,integritee,private_chaincode,pdo,proof_of_luck,rem_pouw,ten_website,yuan2018shadowEth,wang2020hybridchain}. For example, the enclave can write the hashes of the input and output states to the blockchain. Hence, the consistent layer can check that the new state naturally evolves from the latest stored state~\cite{ten_website}. This strategy can protect against rollback attacks if the enclave can always access the latest state in the consistent layer. 
It can also prevent cloning attacks since state updates are tied to the previous state and the hash of the enclave code.

In what follows, we discuss several limitations that affect the three options mentioned above.

\begin{limitation}[Low throughput]
\label{lim:throughput}
    Most permissionless L1 layers exhibit low throughput to ensure safety; for instance, Ethereum has an average block interval of 12 seconds\footnote{\url{https://ycharts.com/indicators/ethereum_average_block_time}, accessed 27.06.2024}, while Bitcoin has an average block interval of 10 minutes\footnote{\url{https://studio.glassnode.com/metrics?a=BTC&category=&m=blockchain.BlockIntervalMean}, accessed 27.06.2024}, severely limiting the number of state updates that can be performed in a certain time interval (when the state is stored in the blockchain) or the granularity of state updates (when the blockchain is used to timestamp responses).
\end{limitation}

\begin{limitation}[Existential honesty]
\label{lim:honesty}
    The only means for enclaves to receive all blockchain history (and keep track of the actual block height or hash) is to (1) either directly participate in consensus~\cite{russinovich2019ccf} or (2) connect to at least one \emph{honest} blockchain node (this is often referred to as the existential honesty assumption~\cite{bitcoin_protocol_analysis}). 
    Running the entire consensus within an enclave~\cite{russinovich2019ccf} is usually discouraged as it increases the TCB size and code complexity,
    thereby increasing the attack surface~\cite{sgx_dev_guide}. Without direct participation in the consensus protocols, 
    enclaves need to connect to multiple blockchain nodes (e.g., lightweight Bitcoin clients are recommended to connect to at least four nodes) to ensure that at least one of their connections is honest and faithfully reports the current block. 
    However, this limitation can be mitigated if (1) enclaves tie their response to their local timestamp (e.g., block height, hash), and (2) clients check whether their timestamp matches the one in the enclave's response.
    In this sense, one can effectively outsource the existential honesty limitation from the enclaves to the clients, who now have to be connected to at least one honest blockchain node to track the current state.
\end{limitation}

\begin{limitation}[Blockchain forks]
\label{lim:forks}
    Permissionless systems like Ethereum and Bitcoin only offer eventual consistency; forks (i.e., blocks with the same height) can naturally occur, which, in turn, would weaken the security provisions of this approach to protect TEE states from forking.
    For example, BITE~\cite{bite} uses enclaves that scan Bitcoin blocks and answer client requests. Since Bitcoin only provides eventual consensus,  assume that a fork occurs in Bitcoin at height $h$ and a light-client $LC$ submitted a transaction $t$ that is only included in one of the forks. Next, $LC$ queries the enclave to determine its balance. A malicious operator can create two clones of the BITE enclave, providing each instance with one of the blockchain forks. $LC$ will receive different balances depending on which clone serves the request. 
    Similarly, Narrator~\cite{Narrator} is a TEE-based anti-forking solution for TEE applications; here, a set of Narrator enclaves provide an anti-forking mechanism for enclaves running arbitrary applications. The security of Narrator holds as long as (1) the set of Narrator enclaves is not forked, and (2) each platform runs at most one Narrator enclave. Towards this end, at start time, a Narrator enclave writes a platform-bound ID to a blockchain; this makes it possible to distinguish if two Narrator enclaves are running on the same machine. Clearly, a fork on the blockchain would allow a malicious operator to run two Narrator enclaves on the same machine. An effective countermeasure to address this problem would be to couple enclave responses with the block height \emph{and} the block hash. This allows clients to determine that this state was computed from a fork.
\end{limitation}

\begin{limitation}[Randomized computations]
\label{lim:random}
There exists another attack avenue on enclaves that execute randomized contracts. In particular, the adversary can run the enclave multiple times---by re-running a single instance repeatedly or by running multiple instances at once---to obtain different outputs, each dependent on the randomness drawn by the enclave instance during the computation. Obtaining different outputs could provide an unfair advantage for the adversary. For example, consider the case where the smart contract outputs a winning lottery ticket. In that case, the adversary can obtain multiple tickets---each output by one of the enclave instances---and decide which one to broadcast as the winning one. Examples of randomized smart contracts that are vulnerable to such attacks include PoUW~\cite{rem_pouw} (cf. Section~\ref{sec:s_stateless}), Proof of Luck~\cite{proof_of_luck}, lottery contracts~\cite{fastkitten} (cf. Section~\ref{sec:s_permissioned}), and Ten~\cite{ten_website} (cf. Section~\ref{sec:cs_ten}). 
An effective solution is to use ephemeral IDs. In particular, if each enclave creates an ephemeral key pair $sk,pk$ at startup and uses $sk$ to sign its output, the adversary cannot obtain multiple outputs that verify with respect to $pk$. This solution works as long as ephemeral enclave IDs (i.e., their public keys) are appropriately managed. For example, in the lottery application described above, clients must agree on the enclave instance (and its public key) that is entitled to draw the winning ticket.
\end{limitation}

\noindent\textbf{A note on timestamping and monotonic counters.} The idea of using the current block height to tell if the TEE state is fresh is reminiscent of ``monotonic counters''. Monotonic counters have been proposed (and used) in the context of TEEs to prevent rollback attacks. That is, an enclave can use a monotonic counter to prevent its local state from being rolled back---e.g., once the enclave has processed a transaction $tx$, the adversary cannot roll back the enclave to a previous state where $tx$ had not been processed. For instance, Milutinovic \emph{et al.}~\cite{proof_of_luck} propose monotonic counters to prevent cloning attacks on TEE-based leader elections. Here, an enclave sleeps for a random period and generates a signed \emph{Proof of Luck (PoL)} afterward, which the miner includes in a block proposal. The PoL protocol increments the monotonic counters on a platform before drawing a random number. The enclave validates that the monotonic counters have the expected value before generating the PoL in the last step of the protocol. If an adversary runs multiple PoL enclaves, each enclave will increase the monotonic counter at the beginning of the protocol, and all but one enclave will see a counter mismatch at the end of the protocol. However, a monotonic counter does not help the enclave distinguish whether it has processed all transactions committed to the blockchain. Furthermore, monotonic counters are usually implemented via TPM registers. As such, they represent a single point of failure and tend to wear out~\cite{matetic2017rote}. In practice, we note that monotonic counters are not viable solutions against rollback attacks since most platforms no longer support them~\cite{sgx_mc_1,sgx_mc_2}.

\vspace{0.5em} \noindent\textbf{Example: CCF.}
Howard \emph{et al.}~\cite{russinovich2019ccf,howard2023ccf} propose a framework for permissioned confidential blockchains (CCF). CCF runs PBFT or RAFT within the enclaves as the underlying consensus protocols (permissioned consensus protocols).  
CCF services are based on a replicated key-value store (KVS). To ensure consistency of the KVS, CCF divides time into a series of views $v$. In each view, one of the nodes is elected as primary while the other nodes serve as backups. Only the primary enclave $E_P$ processes client transactions $tx$ and appends them to the ledger. The current state of the ledger, i.e., all committed transactions, is represented by a Merkle tree. $E_P$ appends every executed transaction to the Merkle tree, regularly signs the Merkle root of the ledger, and appends it to the ledger as a \emph{signature transaction}. A transaction $tx$ is only considered committed if a signature transaction, including $tx$, is replicated on most nodes. 
When the enclave restarts, it recovers the global state by replaying all transactions from the ledger.
When a client connects to CCF, it receives the current view $v_i$, which denotes the current progress in the ledger. Here, $v_i$ serves as a logical timestamp associated with the enclave state, which the client can check before submitting a transaction to the service.

\begin{figure}[]
    \centering
    \scalebox{0.75}{
        \includegraphics[]{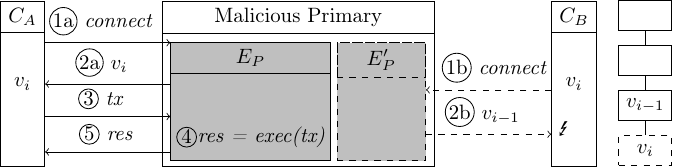}
    }
    \caption{Example of a (failed) cloning attack on CCF~\protect\cite{russinovich2019ccf}. A malicious primary runs two enclaves, $E_P$ and $E'_P$, where $E'_P$ keeps an outdated state. $C_B$ detects a view mismatch and terminates the connection.}
    \label{fig:cloning_ccf}
\end{figure}

Figure~\ref{fig:cloning_ccf} shows how CCF leverages the $v_i$ to prevent cloning attacks. Assume a malicious primary who runs two enclaves $E_P$ and $E'_P$. It provisions $E_P$ and $E'_P$ with all transactions up to the views $v_{i}$ and $v_{i-1}$, respectively. When a client connects to the enclave (step \circled{1}), it first returns its current view (step \circled{2}). The client checks the provided view against the view it knows. Client $C_B$, who is connected to $E'_P$, will detect a mismatch in the views and terminate the connection. Only client $C_A$, connected to $E_P$, sends the transaction $tx$ as the views match (step \circled{3}). The enclave executes the transaction, computing the result (step \circled{4}), and finally returns it to the client (step \circled{5}). Note that a rollback attack would result in an outdated view, so the client would not provide its transaction to the enclave.

Note that CCF overcomes Limitations~\ref{lim:throughput}-\ref{lim:random} as follows. First, CCF's consensus algorithm is based on PBFT and RAFT, which provide high throughput and overcome Limitation~\ref{lim:throughput}. These consensus protocols also guarantee finality, alleviating Limitation~\ref{lim:forks}. CCF executes the whole consensus algorithm in an enclave. Enclaves have a direct view of the block height (i.e., \emph{view}) and do not need to rely on the honesty of other nodes, thus bypassing Limitation~\ref{lim:honesty}. We note, however, that implementing consensus inside the enclave significantly increases the TCB and is not a recommended design choice~\cite{sgx_dev_guide}. Finally, since clients directly connect to the CCF enclave through an encrypted TLS session, a malicious primary cannot inspect different outcomes of a randomized contract execution, alleviating Limitation~\ref{lim:random}. 

\begin{solution}[Serializing State]
\label{sol:serializing}
    Serializing the enclave output using a consistent layer (e.g., the consensus layer of blockchains) can prevent rollback and cloning attacks. However, it needs to be combined with ephemeral IDs to prevent cloning attacks when the TEE computations are non-deterministic. 
\end{solution}

\newcommand{\yes}{\ding{51}}
\newcommand{\nap}{ }
\newcommand{\noo}{\ding{55}}
\newcommand{\stateless}{\yes}
\newcommand{\ephemeral}{\yes}
\newcommand{\fixedSet}{\yes}
\newcommand{\txReplay}{\yes}
\newcommand{\timestamp}{\yes}
\newcommand{\chainState}{\yes}

\newcommand{\resultcolumnA}[9]{ #1 & #2 & #3 & #4 & #5 & #6 & #7 & #8 & #9 & }
\newcommand{\resultcolumnB}[8]{ #1 & #4 & #8 & #2 & #3 & #6 & #7 & #5 \\ \hline}
\begin{table*}[ht]
\centering
\caption{Summary of our analysis of 24 TEE-based blockchain applications from~\cite{awesome_sgx,li2022sokteeassistedconfidentialsmart}. We report for each application which forking mitigation(s) are used and whether they overcome the corresponding limitations. We denote that a countermeasure is used with \yes. Further, we write \noo~(resp. \yes) if the application overcomes a limitation and leave the field empty if a limitation is not applicable because the application does not deploy the corresponding countermeasure.
$\star$ means that the underlying blockchain is not specified in enough detail to reason about Limitations~\ref{lim:throughput}-\ref{lim:random}, respectively.}
\label{table:overview}
\scalebox{0.92}{
    \begin{tabular}
    {|l|HHP{0.85cm}|P{1.0cm}|P{1.2cm}|P{1.1cm}|P{0.9cm}|P{1.2cm}|c|c|c|c|c|c|c|c|c|} 
 \hline
 \multirow{3}{*}{Project} & & & \multicolumn{6}{c|}{Forking Mitigations} & \multicolumn{8}{c|}{Limitations} \\\cline{2-17}
 & \multirow{2}{*}{Rollback} & \multirow{2}{*}{Cloning} & Stateless & Ephemeral & Fixed set & Transaction & Time- & State on & \multicolumn{3}{c|}{Functionality} & \multicolumn{4}{c|}{Robustness} & \multicolumn{1}{c|}{Performance} \\ \cline{10-17}
 & & & enclaves & identities & of clients & replay & stamping & the ledger &\resultcolumnB{L1}{L2}{L3}{L4}{L5}{L6}{L7}{L8}
 \multicolumn{17}{|c|}{\textbf{TEE-based Smart Contracts}} \\ \hline
 \resultcolumnA{Azure CCF~\cite{russinovich2019ccf} }{\yes}{\yes}{\stateless}{ }{ }{\txReplay}{\timestamp}{ }\resultcolumnB{\noo}{\nap}{\nap}{\nap}{\noo}{\noo}{\noo}{\noo}
 \resultcolumnA{CONFIDE~\cite{yan2020confide} }{ }{ }{ }{ }{ }{ }{\timestamp}{ }\resultcolumnB{\nap}{\nap}{\nap}{\nap}{\noo}{\noo}{\noo}{\noo}
 \resultcolumnA{CreDB~\cite{credb} }{\yes}{\yes}{ }{ }{ }{ }{\timestamp}{ }\resultcolumnB{\nap}{\nap}{\nap}{\nap}{$\star$}{$\star$}{$\star$}{$\star$}
 \resultcolumnA{Ekiden~\cite{ekiden} }{\yes}{\yes}{ }{ }{ }{ }{ }{\chainState}\resultcolumnB{\nap}{\nap}{\nap}{\nap}{\noo}{\noo}{\noo}{\noo}
 \resultcolumnA{Phala~\cite{whitepaper_phala} }{\yes}{\yes}{\stateless}{ }{ }{\txReplay}{\timestamp}{ }\resultcolumnB{\noo}{\nap}{\nap}{\nap}{\noo}{\noo}{\yes}{\noo}
 \resultcolumnA{Secret Network~\cite{graypaper_secret_network} }{\noo}{\noo}{\stateless}{ }{ }{\txReplay}{ }{ }\resultcolumnB{\noo}{\nap}{\nap}{\nap}{\noo}{\yes}{\noo}{\noo}
 
 \multicolumn{17}{|c|}{\textbf{TEE-based Consensus Protocols}} \\ \hline
 \resultcolumnA{Crust sWorker~\cite{crust_sworker} }{\yes}{\yes}{ }{ }{ }{ }{\timestamp}{ }\resultcolumnB{\nap}{\nap}{\nap}{\nap}{\noo}{\noo}{\noo}{\noo}
 \resultcolumnA{ENGRAFT~\cite{engraft} }{\yes}{\yes}{\stateless}{\ephemeral}{ }{ }{ }{\chainState}\resultcolumnB{\noo}{\yes}{\nap}{\nap}{\noo}{\noo}{\noo}{\noo}
 \resultcolumnA{MobileCoin~\cite{mobilecoin} }{\yes}{\yes}{\stateless}{ }{ }{ }{ }{ }\resultcolumnB{\noo}{\nap}{\nap}{\nap}{\nap}{\nap}{\nap}{\nap}
 \resultcolumnA{Proof of Luck~\cite{proof_of_luck} }{\yes}{\yes}{\stateless}{ }{ }{ }{ }{\chainState}\resultcolumnB{\noo}{\nap}{\nap}{\nap}{\noo}{\noo}{\noo}{\yes}
 \resultcolumnA{REM~\cite{rem_pouw} }{\yes}{\noo}{\stateless}{ }{ }{ }{ }{\chainState}\resultcolumnB{\noo}{\nap}{\nap}{\nap}{\noo}{\noo}{\noo}{\yes}

 \multicolumn{17}{|c|}{\textbf{TEE-based Layer 2 Solutions}} \\ \hline
 \resultcolumnA{COMMITEE~\cite{erwig2023commitee} }{ }{ }{\yes}{\yes}{ }{ }{ }{ }\resultcolumnB{\noo}{\yes}{ }{ }{ }{ }{ }{ }
 \resultcolumnA{FastKitten~\cite{fastkitten} }{\yes}{\noo}{ }{ }{\fixedSet}{ }{ }{ }\resultcolumnB{\nap}{\nap}{\yes}{\yes}{\nap}{\nap}{\nap}{\nap}
 \resultcolumnA{Hybridchain~\cite{wang2020hybridchain}}{ }{ }{ }{\ephemeral}{ }{ }{ }{\chainState}\resultcolumnB{\nap}{\yes}{\nap}{\nap}{\noo}{\noo}{\noo}{\noo}
 \resultcolumnA{IntegriTEE~\cite{integritee_litepaper} }{\yes}{\yes}{ }{ }{ }{ }{ }{\chainState}\resultcolumnB{\nap}{\nap}{\nap}{\nap}{\noo}{\noo}{\yes}{\yes}
 \resultcolumnA{Obscuro Mixer~\cite{obscuro_mixer} }{\yes}{\yes}{\stateless}{\ephemeral}{ }{ }{ }{ }\resultcolumnB{\noo}{\noo}{\nap}{\nap}{\nap}{\nap}{\nap}{\nap}
 \resultcolumnA{PrivacyGuard~\cite{xiao2020privacyGuard}}{ }{ }{\stateless}{ }{ }{ }{ }{ }\resultcolumnB{\noo}{\nap}{\nap}{\nap}{\nap}{\nap}{\nap}{\nap}
 \resultcolumnA{Private Chaincode~\cite{private_chaincode} }{\yes}{\yes}{ }{ }{ }{ }{ }{\chainState}\resultcolumnB{\nap}{\nap}{\nap}{\nap}{\noo}{\noo}{\noo}{\yes}
 \resultcolumnA{Private Data Objects~\cite{pdo} }{\yes}{\yes}{\stateless}{ }{ }{ }{ }{\chainState}\resultcolumnB{\noo}{\nap}{\nap}{\nap}{\noo}{\noo}{\noo}{\noo}
 \resultcolumnA{ShadowEth~\cite{yuan2018shadowEth}}{ }{ }{ }{ }{ }{ }{\timestamp}{\chainState}\resultcolumnB{\nap}{\nap}{\nap}{\nap}{\noo}{\noo}{\noo}{\noo}
 \resultcolumnA{Teechain~\cite{lind2019teechain} }{\yes}{\yes}{ }{ }{\fixedSet}{ }{ }{ }\resultcolumnB{\nap}{\nap}{\yes}{\yes}{\nap}{\nap}{\nap}{\nap}
 \resultcolumnA{Ten~\cite{ten_website} }{\yes}{\noo}{ }{ }{ }{ }{ }{\chainState}\resultcolumnB{\nap}{\nap}{\nap}{\nap}{\noo}{\noo}{\noo}{\yes}
 \resultcolumnA{Tesseract~\cite{bentov2019tesseract} }{\yes}{\yes}{\stateless}{\ephemeral}{ }{\txReplay}{\timestamp}{ }\resultcolumnB{\noo}{\noo}{\nap}{\nap}{\noo}{\noo}{\noo}{\noo}
 \resultcolumnA{Twilight~\cite{twilight} }{\yes}{\yes}{\stateless}{\ephemeral}{ }{ }{ }{ }\resultcolumnB{\noo}{\noo}{\nap}{\nap}{\nap}{\nap}{\nap}{\nap}
 
 \multicolumn{17}{|c|}{\textbf{TEE-based Blockchain Applications}} \\ \hline
 \resultcolumnA{BITE~\cite{bite} }{\yes}{\yes}{ }{ }{ }{ }{\timestamp}{ }\resultcolumnB{\nap}{\nap}{\nap}{\nap}{\noo}{\noo}{\noo}{\noo}
 \resultcolumnA{LSKV~\cite{lskv_talk} }{\yes}{\yes}{\stateless}{ }{ }{\txReplay}{\timestamp}{ }\resultcolumnB{\noo}{\nap}{\nap}{\nap}{\noo}{\noo}{\noo}{\noo}
 \resultcolumnA{sgxwallet~\cite{sgxwallet} }{\yes}{\yes}{\stateless}{ }{ }{ }{ }{ }\resultcolumnB{\noo}{\nap}{\nap}{\nap}{\nap}{\nap}{\nap}{\nap}
 \resultcolumnA{Ternoa Network~\cite{doc_ternoa} }{\yes}{\yes}{ }{\ephemeral}{ }{ }{\timestamp}{ }\resultcolumnB{\noo}{\noo}{\nap}{\nap}{\noo}{\noo}{\noo}{\noo}
 \resultcolumnA{Town Crier~\cite{town_crier} }{\yes}{\yes}{\stateless}{ }{ }{ }{\timestamp}{ }\resultcolumnB{\noo}{\nap}{\nap}{\nap}{\noo}{\noo}{\noo}{\noo}
\end{tabular}}
\end{table*}

\subsection{Summary of Findings}
\label{sec:summary}

\noindent Table~\ref{table:overview} presents the results of our study.
We summarize our findings below.

\begin{itemize}[leftmargin=*]
    \item Out of the 29 TEE-based blockchain platforms we analyzed,  five are vulnerable to forking attacks. In particular, four platforms are vulnerable to cloning attacks, while one is vulnerable to rollback and cloning attacks. In Section~\ref{sec:case_studies}, we focus on three of those five platforms---those that are either fully deployed or offer at least a testnet---and describe the attacks in detail. We stress that none of the attacks we present were known before. 
   
    \item 16 of the platforms that we considered use stateless enclaves (cf. Section~\ref{sec:s_stateless}). 11 of those platforms rely on the untrusted host to provide state information to the enclave~\cite{proof_of_luck,rem_pouw,town_crier,russinovich2019ccf,lskv_talk,whitepaper_phala,graypaper_secret_network,bentov2019tesseract,pdo,engraft,xiao2020privacyGuard}. 

    \item Seven of the analyzed platforms leverage Ephemeral IDs (cf. Section~\ref{sec:s_ephemeral}). Five of these platforms are stateless and secure against forking attacks. Among those platforms, only three platforms~\cite{engraft,wang2020hybridchain,erwig2023commitee} require proper key management (cf. Limitation~\ref{lim:key_management}). The remaining four platforms do not require key management as the application logic is inherently robust to cloning~\cite{obscuro_mixer,doc_ternoa,bentov2019tesseract,twilight}. For example,~\cite{obscuro_mixer} runs a mixer in the enclave, and the adversary has no advantage in running multiple clones.

    \item Two platforms rely on a fixed set of clients or nodes (cf. Section~\ref{sec:s_permissioned}). As both applications are subject to Limitations~\ref{lim:fault} and~\ref{lim:reconfig}, none supports reconfigurations---hence, the set of participating parties is determined at setup time and remains fixed throughout the contract lifetime~\cite{fastkitten,lind2019teechain}.

    \item 21 platforms serialize state with one of the three techniques we discussed in Section~\ref{sec:s_serial}. 
    \begin{itemize}
        \item Five applications replay transactions at enclave restart to recover the state (cf. Option 1). Four out of these applications additionally timestamp enclave responses to queries~\cite{howard2023ccf,lskv_talk,whitepaper_phala,bentov2019tesseract}. The Secret Network~\cite{graypaper_secret_network} also replays transactions but does not rely on additional serialization mechanisms, allowing forking attacks. 
        
        \item 11 enclaves timestamp their responses to queries (cf. Option 2). BITE~\cite{bite} and Tesseract~\cite{bentov2019tesseract} use the latest block hash whereas the other nine applications~\cite{howard2023ccf,lskv_talk,whitepaper_phala,credb,crust_sworker,doc_ternoa,town_crier,yuan2018shadowEth} timestamp by means of the block height. 

        \item Finally, ten applications store state on the blockchain (cf Option 3). Three platforms leverage the enclave to generate new blocks~\cite{proof_of_luck,ten_website,rem_pouw}. The consensus layer only accepts the block if the included hash is the latest block hash. Similarly, ENGRAFT~\cite{engraft} requires replicas to approve that a state update evolves from the latest state. Six applications directly store their state on the blockchain; the consensus layer ensures linearizability of the state (i.e., that a state can only evolve from the previous state~\cite{ekiden,private_chaincode,pdo,integritee,yuan2018shadowEth,wang2020hybridchain}).
        Only ShadowEth~\cite{yuan2018shadowEth} deploys Options 2 and 3 in parallel.
        
        \item All 21 applications leverage a high-throughput L1 or batch state updates for on-chain storage (cf. Limitation~\ref{lim:throughput}). For example, Tesseract~\cite{bentov2019tesseract} updates client balances in memory and only sends settlement transactions to the L1 once per day. Note that CreDB~\cite{credb} does not provide enough details about the underlying blockchain to determine if it is subject to any limitation. 
        
        \item Only one application requires existential honesty (cf. Limitation~\ref{lim:honesty}):  the Secret Network~\cite{graypaper_secret_network} does not validate the order of transactions the host provides. In CCF~\cite{howard2023ccf},
        the enclaves also actively participate in consensus. 
        
        \item 18 platforms can tolerate blockchain forks (cf. Limitation~\ref{lim:forks}). Nine applications rely on a fully consistent consensus layer~\cite{russinovich2019ccf,lskv_talk,engraft,private_chaincode,pdo,graypaper_secret_network,yan2020confide,yuan2018shadowEth,wang2020hybridchain} to serialize events. Five additional applications cryptographically bind the output (including the block hash) on chain~\cite{ekiden,proof_of_luck,rem_pouw,ten_website,doc_ternoa}, and two applications make the current block available to their clients~\cite{bite,bentov2019tesseract}. Additional two applications are not subject to blockchain forks because of their application logic~\cite{crust_sworker,town_crier}.
        
        \item A total of 15 platforms do not allow the exploitation of randomized computations (cf. Limitation~\ref{lim:random}). Eight of these platforms run deterministic contracts~\cite{bite,crust_sworker,ekiden,lskv_talk,doc_ternoa,bentov2019tesseract,town_crier,yan2020confide}. Another seven platforms encrypt the enclave response for the client so a malicious host cannot read it~\cite{howard2023ccf,engraft,whitepaper_phala,pdo,graypaper_secret_network,wang2020hybridchain,yuan2018shadowEth}.
    \end{itemize}
\end{itemize}

\begin{figure}[tb]
	\centering
	\scalebox{0.75}{
		\includegraphics[]{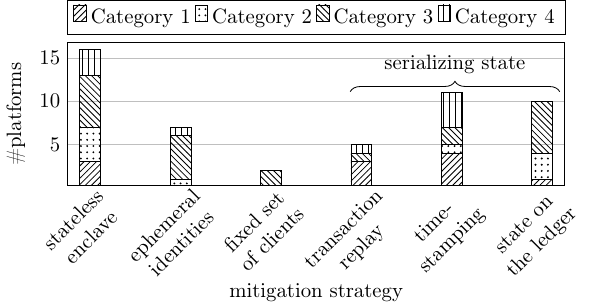}
	}
	\caption{Distribution of the mitigation strategies used by the platforms in Table~\ref{table:overview}. Note that platforms may use more than one strategy to prevent forking.}
	\label{fig:histogram_mitigations}
\end{figure}

\vspace{0.5em}
\noindent While these findings show the adoption of the mitigation strategies across the whole set of analyzed applications, we now additionally analyze the adoption of the mitigation strategies across the different system categories identified in Section~~\ref{sec:sys_classification}. Figure~\ref{fig:histogram_mitigations} depicts the distribution of these mitigations across the different system categories. The following key observations can be made based on our results:

\begin{itemize}
	\item \textbf{Stateless enclaves} are employed by platforms in all four categories.
	\newpage
	\item \textbf{Ephemeral identities} are primarily used by platforms in Category 3 (TEE-based Layer 2 solutions) to counter cloning attacks. Specifically, five out of 13 platforms in Category 3 utilize ephemeral IDs. 
	In contrast, no platform in Category 1 (TEE-based smart contracts) uses ephemeral identities.
	\item \textbf{A fixed set of clients} is a technique exclusively employed by platforms in Category 3.
	\item \textbf{State serialization} techniques are utilized across all four system categories. Platforms in Category 1 mostly rely on transaction replay and timestamping, whereas Category 2 and 3 more commonly store their states on the ledger. Many platforms in Category 4 (TEE-based blockchain applications) use timestamping. Lastly, no platform in Category 2 replays transactions to recover state information, and no enclave in a blockchain application stores its state on the ledger.
\end{itemize}

\noindent\textbf{Summary: } We conclude that mitigations for forking attacks introduce trade-offs in terms of (1) types of applications that can be deployed, (2) tolerance to peers joining/leaving the network, and (3) overall complexity of the platform. 
For example, stateless enclaves can prevent rollback attacks but limit the type of contracts that can be deployed. For contracts that require a local state, rollback attacks can be mitigated if the state is serialized on L1. The latter can be achieved in several ways. However, the choice of the L1 layer is very important as it determines the effectiveness of the serialization technique. For instance, a low-throughput L1 will result in a coarse-grained timestamping mechanism. 
We note that ten platforms that serialize enclave state using the L1 layer use permissioned blockchains to achieve high throughput and finality.

With respect to cloning attacks, deterministic smart contracts are safe as long as the platform guarantees that all smart contracts process the same events and in the same order (e.g., by serializing state on L1). If the smart contract is randomized, ephemeral IDs allow to distinguish clones of the same contract but may require proper key management and means to determine, at any time, the set of enclaves that are part of the platform (and their IDs).

%% file: phala.tex
\begin{figure}[]
    \centering
    \scalebox{0.73}{
        \includegraphics[]{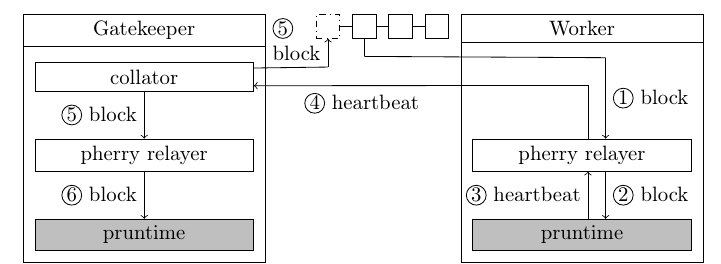}
    }
    \caption{Overview of the components in Phala~\protect\cite{whitepaper_phala}. Worker pruntimes regularly issue heartbeat transactions to the ledger, which are checked by Gatekeeper pruntimes.}
    \label{fig:phala_overview}
\end{figure}

\vspace{0.5em}\noindent \textbf{Overview: }\emph{Phala} \cite{whitepaper_phala} is a Layer 1 (L1) blockchain that is built using the Substrate framework ~\cite{substrate_blockchain_framework}. It acts as a para-chain that plugs into \emph{Polkadot}\cite{polkadot_website}. At the time of writing, Phala has an active mainnet with a market cap of \$90M~\cite{marketcap_phala}, $157$ deployed contracts, and $150$ Worker nodes serving over 800k off-chain queries per day~\cite{phala_number_of_contracts}. Phala leverages TEEs to enable off-chain confidential smart contract execution. 
Figure~\ref{fig:phala_overview} shows the components of the Phala Network and how they interact. There are two types of nodes in the Phala Network: \emph{Gatekeepers} and \emph{Workers}. Both nodes operate a \emph{pruntime} enclave connected to the network through a \emph{pherry relayer}. Gatekeepers additionally run a \emph{collator}, which participates in the consensus.
Phala uses Authority-round (Aura) consensus, a Proof-of-Authority (PoA) leader election algorithm that elects the next leader to create the next block. At the time of writing, the developers control all leaders in the system. 
Workers are responsible for smart contract execution. They are assigned specific smart contracts---typically, a contract is run by a single Worker, but multiple Workers for the same contract can be used. Phala derives an asymmetric \emph{contractKey} for each smart contract using a global \emph{masterKey} accessible only to Gatekeeper enclaves. The Worker enclave queries the \emph{contractKey} from the Gatekeeper, seals it, and checkpoints its contract state to disk to provide fault tolerance. 
When a Worker receives a new block, it parses out all transactions for its contract, decrypts them, and processes them. Further, each Worker has an HTTP endpoint, which clients can use to issue contract queries. Phala incentivizes these contract queries, as they do not cost gas fees and result in fast responses. 

As each smart contract is only executed by one enclave, Phala implements heartbeats as a keep-alive mechanism, shown in Figure~\ref{fig:phala_overview}. In particular, the Worker's pherry relayer regularly fetches new blocks from the blockchain (step~\circled{1}) and forwards them to the pruntime (step~\circled{2}). The pruntime computes a function of blockchain meta-data and its public key hash to determine whether or not to send a heartbeat. 
The function is designed so that approximately 20 Workers send a heartbeat in each block. In other words, a Worker pruntime issues a heartbeat roughly every 45 seconds. A heartbeat is sent to the pherry relayer as a transaction containing the current block height (step~\circled{3}). The pherry relayer submits the heartbeat transaction to a  Gatekeeper's collator (miner) (step~\circled{4}) who verifies the transaction and includes it in the next block (step~\circled{5}).
The Gatekeeper's pherry relayer fetches the block and forwards it to the Gatekeeper pruntime (step~\circled{6}), which extracts, validates, and logs the heartbeat.

\vspace{0.5em}\noindent \textbf{Cloning Attack on Contract Queries:} As shown in Figure~\ref{fig:phala_overview}, Phala enclaves directly receive blocks from the pherry relayers (who fetch them from the ledger). Note that the pherry relayers do not run within the TEEs and, hence, could easily perform rollback attacks on the enclaves by providing a stale state from the ledger. 
Thus,
Phala enclaves issue heartbeat transactions every 45 seconds (cf. Figure~\ref{fig:phala_heartbeat} for the exact format of the heartbeat messages). While heartbeats contain the block height (and, as such, could be used for timestamping), enclaves do not check whether they receive regular heartbeats (or acknowledgments) from others. This check is performed only by the Gatekeeper (cf. Figure~\ref{fig:phala_overview}) and does not reside within the Worker enclave. This allows enclaves to be cloned and even isolated from the rest of the network. 

\begin{figure}[tb]
    \centering
    \scalebox{0.75}{
        \includegraphics[]{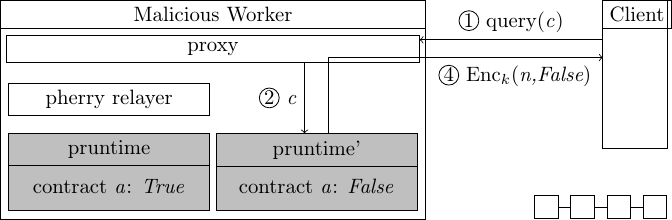} 
    }
    \caption{Sketch of the cloning attack on Phala~\protect\cite{whitepaper_phala}. A malicious Worker clones the pruntime running the smart contract. It then prevents the clone from receiving state updates and answers contract queries with an outdated state.}
    \label{fig:phala_attack_new}
\end{figure}

\begin{figure}[tb]
\centering
\footnotesize
    \begin{lstlisting}[basicstyle=\ttfamily\scriptsize,numbers=left,numberstyle=\tiny\color{gray},xleftmargin=2.5em,frame=single,framexleftmargin=2em,xrightmargin=0.5em]
#[pink::contract(env=PinkEnvironment)]
mod phat_important_data {
    use super::pink;
    use pink::{PinkEnvironment};

    #[ink(storage)] //persistent storage
    pub struct ImportantData {
        data: bool,
    }
    
    impl ImportantData {
        #[ink(constructor)]
        pub fn new() -> Self {
            Self { data: true }
        }
        
        #[ink(message)] //on-chain command 
        pub fn toggle(&mut self) {
            self.data = !self.data;
        }
        
        #[ink(message)] //contract query
        pub fn get_data(&self) -> bool {
            self.data
}}}
    \end{lstlisting}
    \caption{Example of a smart contract in Phala~\cite{whitepaper_phala} that persists storage. Clients can change the stored boolean value via on-chain transactions or read the contract state through a contract query.
    }
    \label{fig:phala_vulnerable_contract}
\end{figure}

We note that contract queries are encrypted with a symmetric key $k$ derived from a key exchange that uses an ephemeral key of the client and the \emph{contractPubKey} of the smart contract. In particular, the client computes $k$ as follows:
\begin{eqnarray*}
    k &=& \mathrm{ECDHKE}(\mathrm{clientPrivKey}, \mathrm{contractPubKey})) \\
       &=& \mathrm{ECDHKE}(\mathrm{clientPubKey}, \mathrm{contractPrivKey}))
\end{eqnarray*}
\noindent where ECDHKE is an Elliptic-curve Diffie–Hellman key exchange and clientPrivKey is the ephemeral private key of the client. The client then computes an encrypted query:
\begin{eqnarray*}
    \mathrm{payload} &=& \mathrm{AEAD\_IV} ||\mathrm{clientPubKey} ||\\
    &&\mathrm{AES}^{GCM}_{k}(\mathrm{contractAddress} || \mathrm{n} || \mathrm{rawQuery})\\
    \mathrm{query} &=& \mathrm{payload} || \mathrm{clientIdentityPubKey} ||\\
    &&\mathrm{sign}_{\mathrm{clientIdentityPrivKey}}(\mathrm{payload})
\end{eqnarray*}
\noindent where AEAD\_IV is an IV for the AES encryption, contractAddress is the address of the queried smart contract, clientIdentityPubKey is a persistent client identity for access control, $n$ is a random nonce reflected in the response, and rawQuery is the actual query.

Given the above setting, an adversary can operate two instances of the pruntime and freely choose the instance to answer a request. The attack is depicted in Figure~\ref{fig:phala_attack_new}. We assume a simple contract with address \emph{a} using a single boolean variable as state, initialized to \emph{False}. At a certain point, a transaction \emph{tx} causes the boolean variable to be set to \emph{True}. From this moment, clients issuing contract queries to the contract at address \emph{a} should receive \emph{True} as a response. However, assume that the adversary creates a clone of the pruntime. The adversary does not start a pherry relayer for the clone, effectively isolating it from the network. The first instance is still connected to the network and issues regular heartbeats. As the cloned pruntime did not receive \emph{tx} (it is isolated from the network), its internal state remains \emph{False}.
At this stage, a client issues a contract query (step~\circled{1}) to the smart contract at address \emph{a}. The adversary forwards the request to the isolated pruntime instance (step~\circled{2}), which decrypts the query and provides \emph{False} as a response to the client (step~\circled{3}).

\vspace{0.5em}\noindent \textbf{Implementation: } 
We implemented and evaluated the attack on a local Phala Testnet version \emph{v2.1.0}~\cite{phala_git_version} using the official Phala docker images. We stress that no real contract was affected while we were validating our attack and that it had no impact whatsoever on the real Phala Network. The adversary operates a machine equipped with an Intel Xeon E-2286G CPU, 128GB of memory, and Ubuntu 22.04.4 LTS. We configure a single node with a Gatekeeper and a Worker pruntime, each connected to a pherry relayer providing new blocks from the blockchain to the enclave.
As for the victim contract, we used a simplified version of the official Phala demo \emph{flip} contract~\cite{phala_flip_contract}, which holds a boolean variable that can be toggled (cf. Figure~\ref{fig:phala_vulnerable_contract}). After initializing the contract to \emph{False}, we start a second Worker pruntime (providing it with the sealed data from the first Worker) and a corresponding pherry relayer. We then terminate the pherry relayer of the second instance, effectively isolating it from the network. We instruct our client to call the deployed smart contract \emph{a}, toggling its state to \emph{True}. The transaction is only processed by the enclave with a connected pherry relayer, such that the isolated enclave remains in state \emph{False}. We then instruct our client to query \emph{a}. At this stage, our proxy intercepts the request and forwards it to the isolated enclave, which returns \emph{False}. Figure~\ref{fig:phala_attack_new} shows that the client cannot distinguish which enclave answered the request as the reply only contains the state itself and the reflected nonce. 

\vspace{0.5em}\noindent \textbf{Suggested Countermeasure: }
Heartbeats in Phala offer a good opportunity to integrate a timestamping mechanism that allows enclaves to self-detect forking attacks: (1) they are authenticated by the enclaves, (2) they contain the block height, and (3) they are sent regularly every 45 seconds. We suggest leveraging those heartbeat messages to ensure that all enclaves are aware of the current block height (and hence the current state). To this end, we suggest that enclaves exchange heartbeats via a separate P2P network and check that they regularly receive heartbeat messages from others (i.e., they are not eclipsed).
A major challenge with this approach lies in Limitation~\ref{lim:honesty} (existential honesty): enclaves need to ensure they are connected to at least one honest node to get the latest state from the network (and reflect it in their heartbeat messages).

This, alone, however, is not sufficient to deter forking attacks. 
Here, we suggest that the enclaves include the latest block height as a timestamp in responses to all contract queries (as suggested in the whitepaper~\cite{whitepaper_phala}). This burdens the requesting client to determine whether the output corresponds to a fresh state and is, therefore, valid. In case Phala opts to support randomized contract execution, we suggest the reliance on ephemeral IDs (cf. Limitation~\ref{lim:random}) as well.

\begin{figure}[tb]
\centering
\footnotesize
    \begin{lstlisting}[basicstyle=\ttfamily\scriptsize,xleftmargin=2.5em,frame=single,framexleftmargin=2.0em,xrightmargin=0.5em,escapeinside={<@}{@>}]
    Heartbeat =  {
        session_id,
        <@\textcolor{blue}{challenge\_block}@>,
        challenge_time,
        iterations,
        n_clusters,
        n_contracts}
    \end{lstlisting}
    \caption{Structure of a heartbeat in Phala~\cite{whitepaper_phala}. The block height is in blue.
    }
    \label{fig:phala_heartbeat}
\end{figure}

%% file: secret.tex
\vspace{0.5em}\noindent \textbf{Overview:} 
The Secret Network \cite{graypaper_secret_network} (SN) is an L1 blockchain that is built using the Cosmos SDK \cite{cosmos_sdk}. At the time of writing, it has a market cap of \$110M~\cite{marketcap_secret} and features $1105$ smart contracts uploaded from $197$ different accounts. 

Each SN node uses a TEE to execute smart contracts.  All cryptographic material used in SN is derived from a \emph{consensus seed} shared among all SN enclaves. To obtain the consensus seed, an enrolling enclave must prove---via remote attestation---to one of the SN enclaves that it is running the genuine SN enclave binary. The consensus seed is sealed to disk to avoid re-enrollment in case an SN enclave is restarted. 

Transactions to invoke a contract are encrypted with a network-wide public key (called \emph{consensusIoPubKey}) and can be decrypted by any SN enclave. Once a transaction is chosen to be included in the next block, it is decrypted and executed inside the enclave; the contract output is encrypted (e.g., with the sender's public key) and included in the block as well. 
SN allows contract queries to read the state of the smart contract. Enclaves, in particular, have an HTTP endpoint that the client can use to request the contract state. This design was chosen to avoid gas fees and reduce delays for contract queries. 

\vspace{0.5em}\noindent \textbf{Cloning Attack on Contract Queries:} Previous work~\cite{sgx_oneratete_paper} has shown that the SN did not feature protection against rollback attacks and that an adversary could learn private data of a transaction (e.g., sender, receiver, amount) by rolling-back the enclave and replaying transactions. In what follows, we present a new forking attack on the Secret Network based on cloning.

We note that contract queries are encrypted with a symmetric key $k$ derived from a key exchange that uses an ephemeral key of the client and the \emph{consensusIoPubKey} of the SN enclave. In particular, the client computes $k$ as follows:
\begin{eqnarray*}
    k' &=& \mathrm{ECDHKE}(\mathrm{clientPrivKey}, \mathrm{consensusIoPubKey}) \\
       &=& \mathrm{ECDHKE}(\mathrm{clientPubKey}, \mathrm{consensusIoPrivKey}) \\
    k &=& \mathrm{HKDF}(n,k')
\end{eqnarray*}
\noindent where $\mathrm{ECDHKE}$ is an Elliptic-curve Diffie–Hellman key exchange, $\mathrm{HKDF}$ is a key derivation function, $\mathrm{clientPrivKey}$ is the ephemeral private key of the client, and $n$ is a nonce chosen by the client. Note that $n$ and $clientPubKey$ (i.e., the client's ephemeral public key) are sent in cleartext in the transaction to enable the SN enclave to derive the same symmetric key $k$.
We point out that other fields of a transaction are not encrypted---e.g.,
the contract address ($contractAddress$) is sent as cleartext. Hence, a query has the following format:
\begin{eqnarray*}
    \mathrm{query} &=& \mathrm{contractAddress} || \mathrm{n} || \mathrm{clientPubKey} ||\\
    &&\mathrm{AES}^{SIV}_{k}(\mathrm{codeHash} || \mathrm{rawQuery})\\
\end{eqnarray*}
\noindent where $\mathrm{codeHash}$ is the hash of the contract that should handle the query, and $\mathrm{rawQuery}$ is the actual query.

\begin{figure}[]
    \centering
    \scalebox{0.75}{
        \includegraphics[]{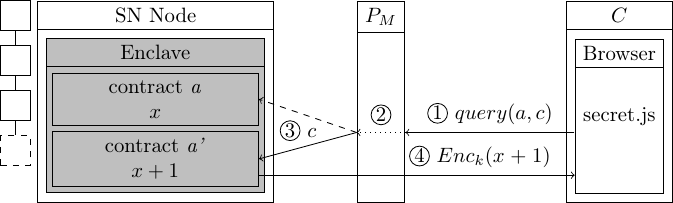}
    }
    \caption{Sketch of the cloning attack on the Secret Network~\protect\cite{graypaper_secret_network}. A malicious Proxy $P_M$ in the network changes the contract address in the client's query to return the state of a different instance with the same code. }
    \label{fig:secret_attack_mitm_new}
\end{figure}

Given the above format, an adversary can simply change the $\mathrm{contractAddress}$ field of a query and use another instance of the same contract (matching the $\mathrm{codeHash}$ in the query) on any network node to answer the query.
The attack is depicted in Figure~\ref{fig:secret_attack_mitm_new}. We assume a simple contract with address \emph{a} that uses a single counter variable as state, initialized to $\mathrm{x}$. At a certain point, a transaction $tx$ causes the variable to be incremented to $\mathrm{x+1}$. From this moment, clients issuing contract queries to the contract at address  \emph{a} should receive $\mathrm{x+1}$ as a response.
However, assume that the adversary creates a clone of the contract and assigns it to address \emph{a'}. As the contract at address \emph{a'} did not receive $tx$ (it is a different contract instance), its internal state remains $\mathrm{x}$. Note that the contract enclaves at address \emph{a} and at address \emph{a'} share the same $\mathrm{codeHash}$.
At this stage, a client issues a contract query (step \circled{1}) for the enclave at address  \emph{a}. The adversary intercepts the HTTP request (step \circled{2}) and changes the contract address in the requested URL to \emph{a'} (step \circled{3}). Hence, the contract enclave at address \emph{a'} decrypts the query and provides $\mathrm{x}$ as a response to the client (step \circled{4}).

\vspace{0.5em}\noindent \textbf{Implementation: } 
We implemented and evaluated the attack on a Secret Network Testnet version  \emph{v1.13.1}~\cite{secret_git_version}. We stress that no real contract was affected while we were validating our attack and that it had no impact whatsoever on the real Secret Network. The adversary operates a machine equipped with an Intel Xeon E-2286G CPU, 128GB of memory, and Ubuntu 22.04.4 LTS. The victim uses the Firefox browser version 127.0.2. We configure the Secret Network lightweight client \emph{secretcli}~\cite{secretcli} to interact with the testnet. 
As for the victim contract, we used the official Secret Network contract template~\cite{secret_contract_template}, which implements a simple counter (cf.  Figure \ref{fig:secret_vulnerable_contract}). We uploaded the contract to the testnet and launched two clone instances of the contract. During the setup, we specified an initial value $1$ for the counter on each clone, respectively. We now have two clones of the same smart contract: a contract $C_a$ at address $a$ and another $C_{a'}$  at address $a'$, both with counter value $1$. Both contracts are executed by every SN node. We issued a transaction that increments the value of $C_a$ to $2$.  Note that the transaction includes $a$ (as it targets $C_a$), so the counter of $C_{a'}$ stays in state $1$ although both contracts run in the same enclave. We then built a simple website that uses the official SN lightweight client library \emph{secret.js}~\cite{secretjs} to query the state of $C_a$. Queries are handled via HTTP, so we instantiate a malicious HTTP proxy between our client and the testnet. We let a first query to $C_a$ reach the intended contract so that it returns $2$. We then instruct our client to query $C_a$ again. At this stage, the proxy $P_M$ intercepts the HTTP request and replaces address $a$ with $a'$ before feeding it to the enclave, which now returns $1$. Figure~\ref{fig:secret_attack_mitm_new} shows that the client cannot distinguish if the query was answered by $C_a$ or $C_{a'}$ as the reply only contains the state itself. 

\vspace{0.5em}\noindent \textbf{Suggested Countermeasures: } 
We note that an SN contract is assigned an ID (contract address), but the ID is not bound to the messages exchanged with clients. We note that SN IDs are instance-specific; in other words, two contract clones (same binary, same machine) will get different IDs (similar to the ephemeral IDs we describe in Section~\ref{sec:s_ephemeral}). 
A straightforward fix to deter cloning attacks would be to cryptographically bind the contract ID to the client request by including it in the encrypted request payload so that the contract instance can tell if it is the intended receiver.

However, the solution just described does not mitigate rollback attacks. Jean-Louis \emph{et al.}~\cite{sgx_oneratete_paper} suggest implementing a proof-of-publication to ensure that transactions have been committed and ordered on-chain before executing them. This effectively serializes transactions (cf. Takeaway~\ref{sol:serializing}). The Secret Network leverages Tendermint, a BFT version of delegated Proof-of-Stake providing some form of finality (cf. Limitation~\ref{lim:forks}) and relatively short block generation times with a throughput of 10000 transactions per second~\cite{secret_website} (cf. Limitation~\ref{lim:throughput}). However, such an approach would still be limited by the existential honesty assumption (cf. Limitation~\ref{lim:honesty}) as the enclave needs to be connected to at least one honest consensus node to ensure 
access to the latest transactions to be up to date. This is particularly worrisome as Tendermint does not offer strong protection against so-called long-range attacks, where the ledger's history can be rewritten from a past point in time (somewhat analogously to rollback attacks)~\cite{long-range_attacks}. A more elegant alternative to deal with rollback attacks against the SN would be to rely on the TEEs to track the set of TEEs (and their ephemeral IDs) that are members of the network (cf. Takeaway~\ref{sol:eid}). This countermeasure, combined with 
proper reliance on ephemeral IDs as we suggest, would offer a comprehensive solution for the Secret Network against forking attacks. 

\begin{figure}[tb]
\centering
\footnotesize
    \begin{lstlisting}[basicstyle=\ttfamily\scriptsize,numbers=left,numberstyle=\tiny\color{gray},xleftmargin=2.5em,frame=single,framexleftmargin=2em,xrightmargin=0.5em]
pub struct InstantiateMsg { pub count: i32,}
pub enum ExecuteMsg { Increment { }, }
pub enum QueryMsg { GetCount { }, }
pub struct CountResponse { pub count: i32, }

#[entry_point] // constructor
pub fn instantiate(deps: DepsMut, _env: Env,
    info: MessageInfo, msg: InstantiateMsg,
)->StdResult<Response> {
    let state = State {
        count: msg.count,
        owner: info.sender.clone(),};
    config(deps.storage).save(&state)?;
    Ok(Response::default())
}

#[entry_point] // on-chain transaction
pub fn execute(deps: DepsMut, env: Env,
    info: MessageInfo, msg: ExecuteMsg
)->StdResult<Response> {
    match msg { ExecuteMsg::Increment { } => 
			try_increment(deps, env), }
}

pub fn try_increment(deps: DepsMut, _env: Env
)->StdResult<Response> {
    config(deps.storage).update(|mut state| -> 
	Result<_,StdError>{state.count+= 1;Ok(state)})?;
    Ok(Response::default())
}

#[entry_point] // contract query
pub fn query(deps: Deps, _env: Env, msg: QueryMsg
)->StdResult<Binary> {
    match msg { QueryMsg::GetCount { } => 
	to_binary(&query_count(deps)?), }
}

fn query_count(deps: Deps)->StdResult<CountResponse> {
    let state = config_read(deps.storage).load()?;
    Ok(CountResponse { count: state.count })
}
    \end{lstlisting}
    \caption{Vulnerable smart contract in the Secret Network~\protect\cite{secret_contract_template} that persists storage. Clients can increase the stored integer value via on-chain transactions or read the contract state through a contract query.
    }
    \label{fig:secret_vulnerable_contract}
\end{figure}

%% file: ten.tex
\vspace{0.5em}\noindent \textbf{Overview:} 
Ten~\cite{ten_website} (formerly \emph{Obscuro}) is an L2 solution built on top of Ethereum~\cite{ethereum_whitepaper}. At the time of writing, Ten has a public testnet; the mainnet is expected to go live in Q3 2024. Each Ten node uses a TEE to execute transactions in a privacy-preserving manner. By running the Ethereum Virtual Machine (EVM) inside the TEE, any Ethereum smart contract can be ported to Ten. As mentioned earlier, all cryptographic material in Ten is derived from a \emph{master seed} shared among all Ten enclaves. At enrollment, an enclave can obtain the master seed by providing its attestation report. Any Ten enclave can verify this attestation and, when successful, can encrypt the master seed with the enrolling enclave's public key for provisioning. The enclave seals cryptographic key material and smart contract state to avoid re-enrollment when the enclave is restarted. 
The enclave stores its data, including the \emph{master seed} in \emph{EdgelessDB}\cite{edgelessdb}, a TEE-based SQL database that seals all its data to disk to provide fault tolerance.

Transactions are encrypted with a network-wide public key (called \emph{networkKey}) and can be decrypted by any Ten enclave. Transactions are ordered in the context of \emph{rollups} through a custom \emph{Proof-of-Block-Inclusion (POBI)} consensus protocol. In particular, the enclave extracts the previous rollup from the latest L1 block and generates the next rollup on top of it, including a random nonce. The rollup with the lowest nonce is committed to the L1 layer via a dedicated Ethereum smart contract. Rollup generation uses a throttling mechanism based on proof-of-work to ensure that any Ten enclave can output at most one rollup with its corresponding nonce per block. 
Ten enclaves are not vulnerable to rollback attacks. This is because rollups are bound to the current L1 block. If the Ten enclave is rolled back, it will output a rollup bound to a stale L1 block; then, L1 will treat the rollup as invalid and discard it.

\begin{figure}[]
    \centering
    \scalebox{0.75}{
        \includegraphics[]{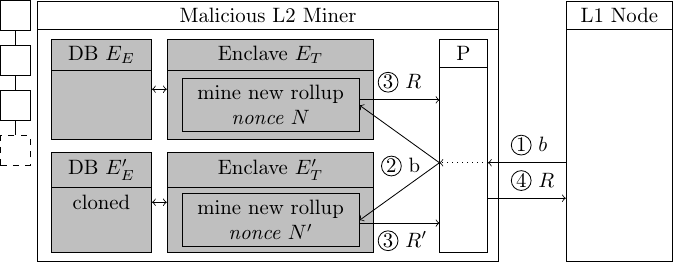}
    }
    \caption{Sketch of the cloning attack on Ten~\protect\cite{ten_website}. An adversary increases the chances of proposing the next block by running two enclave clones and choosing the output with the lowest nonce.}
    \label{fig:ten_attack_overview_new}
\end{figure}

\vspace{0.5em}\noindent \textbf{Cloning Attack on Block Generation:} Despite being rollback-resistant, we show that Ten enclaves are vulnerable to cloning attacks that target the POBI consensus protocol. In a nutshell, POBI uses Ten 
enclaves to draw a random nonce, and the enclave with the lowest nonce is allowed to propose the next rollup. By cloning the enclave, an adversary can increase the chances that one of its enclaves is allowed to propose the next rollup.
Ten attempts to combat cloning attacks by requiring a registration fee for parties to enroll their enclaves. In principle, this requires clients to pay the enrollment fee $n$ times when enrolling $n$ enclaves to increase their chances of proposing the next rollup. However, an adversary can circumvent this measure and clone the Ten enclave \emph{after} enrollment. That is, the adversary can create $n$ clones of the Ten enclave while paying the enrollment fee only for one. Note also that this attack is effective, despite the throttling mechanism used by Ten (enclaves must do some variant of proof of work by computing a large number of hashes at restart time)
to ensure that each enclave proposes at most one rollup per round.
Our attack is depicted in Figure~\ref{fig:ten_attack_overview_new} and works as follows. The adversary starts a Ten enclave that completes enrollment, obtains the master seed, and seals cryptographic keys to disk. Hence, the adversary creates a clone of the enclave on the same machine. At this time, both enclave instances have access to the state sealed by the first enclave. Next, the adversary receives a new block from an L1 node (step~\circled{1}) and feeds it to both clones (step~\circled{2}). The enclaves generate random nonces $N$ and $N'$, respectively, and include them in the proposed rollup (step~\circled{3}). The adversary selects the rollup with the lowest nonce and submits it to the L1 layer (step~\circled{4}). 

\vspace{0.5em}\noindent \textbf{Implementation:}
We implemented and evaluated our attack on a local Ten Testnet version \emph{v0.24.7}~\cite{ten_git_version}. In our setup, the adversary operates a machine equipped with an Intel Pentium Silver J5040 CPU, 32GB of memory, and Ubuntu 22.04.03 LTS. Here, we had to register a Ten node and instantiate an EdgelessDB instance~\cite{edgelessdb} (each running in a docker container) to interact with the testnet. 
We then cloned the Ten enclave and connected each enclave to a cloned instance of the EdgelessDB to recover the state. We now have two operational Ten enclaves $E_{T}$ and $E'_{T}$ on the same node. We then use a simple proxy \emph{P} that handles incoming L1 blocks and feeds them to both Ten enclaves. The enclaves generate the random nonces $N$ and $N'$ and include them in the rollups $R$ and $R'$, respectively. The proxy \emph{P} retrieves $R$ and $R'$ from the enclaves and sends the more favorable rollup back to the L1 layer.\footnote{Note that the current implementation of the Ten enclaves is incomplete and does not return the nonces.} 

\vspace{0.5em}\noindent \textbf{Suggested Countermeasures:} 
Ten implements stateful enclaves (to seal the state of the rollups in an Edgeless DB backend). It incorporates a rollback detection mechanism by serializing state. Here, the state is serialized with the block hash seen by the enclave. If the block hash is stale, the L1 layer will not accept the commit request from the enclave. However, as discussed in Section~\ref{sec:s_stateless}, such stateful solutions should also incorporate mechanisms to prevent cloning. An effective anti-cloning mechanism in this particular case would be to rely on ephemeral IDs (cf. Takeaway~\ref{sol:eid}) specific to each enclave. In particular, the L1 contract handling rollups can be easily modified to keep track of the (ephemeral) identities of the TEE enclaves. For instance, the rollup header can include a new field \emph{AggregatorEphemeralID}. Figure~\ref{fig:ten_rollup_header} shows the rollup header of the current Ten implementation and our suggested modification to deter cloning attacks. As we discuss in Section~\ref{sec:s_ephemeral}, this would deter cloning attacks.

\begin{figure}[tb]
\centering
\footnotesize
    \begin{lstlisting}[basicstyle=\ttfamily\scriptsize,xleftmargin=2.5em,frame=single,framexleftmargin=2.0em,xrightmargin=0.5em,escapeinside={<@}{@>}]
    RollupHeader =  {
        L1BlockHeader,
        CrossChainMessages,
        PayloadHash,
        PayloadHashSignature,
        BatchSeqNum,    
        <@\textcolor{blue}{AggregatorNonce}@>,
        <@\textcolor{blue}{AggregatorL2Address}@>,
        <@\textcolor{red}{AggregatorEphemeralID}@>}
    \end{lstlisting}
    \caption{Rollup header of Ten~\protect\cite{ten_website}. Additional fields based on~\cite{whitepaper_obscuro} are shown in blue; our recommendation to incorporate ephemeral IDs is shown in red.
    }
    \label{fig:ten_rollup_header}
\end{figure}